\documentclass[journal,10pt,twoside]{IEEEtran} 
\pdfoutput=1
\usepackage{amsmath,cite,amsfonts,amssymb,psfrag,amsthm,paralist}
\usepackage{graphicx}
\usepackage{epstopdf}
\usepackage{rotating}
\usepackage{dsfont}
\usepackage{color}
\usepackage{tikz}
\usetikzlibrary{plotmarks}
\usepackage{pgfplots}
\usetikzlibrary{calc}
\usetikzlibrary{shapes,arrows}
\usetikzlibrary{decorations.markings}
\usetikzlibrary{positioning}
\pgfplotsset{compat=1.10}
\usetikzlibrary{calc}
\usetikzlibrary{shapes,arrows}
\usetikzlibrary{decorations.markings}

\usepackage[utf8]{inputenc}
\usepackage{authblk}

\usepackage{balance}

\DeclareFontFamily{U}{mathx}{\hyphenchar\font45}
\DeclareFontShape{U}{mathx}{m}{n}{<-> mathx10}{}
\DeclareSymbolFont{mathx}{U}{mathx}{m}{n}
\DeclareMathAccent{\widebar}{0}{mathx}{"73}

\def\BibTeX{{\rm B\kern-.05em{\sc i\kern-.025em b}\kern-.08em T\kern-.1667em\lower.7ex\hbox{E}\kern-.125emX}}

\DeclareFontFamily{U}{mathx}{\hyphenchar\font45}
\DeclareFontShape{U}{mathx}{m}{n}{<-> mathx10}{}
\DeclareSymbolFont{mathx}{U}{mathx}{m}{n}

\newcommand{\Enc}{\mathsf{Enc}}
\newcommand{\Dec}{\mathsf{Dec}}
\newcommand{\Est}{\mathsf{Est}}

\makeatletter
\newtheorem*{rep@theorem}{\rep@title}
\newcommand{\newreptheorem}[2]{%
  \newenvironment{rep#1}[1]{%
    \def\rep@title{\Cref{##1}}%
    \begin{rep@theorem}}%
    {\end{rep@theorem}}}
\newcommand*{\textlabel}[2]{%
  \edef\@currentlabel{#1}
  \phantomsection
  #1\label{#2}
}
\makeatother

\usepackage{scalerel}
\usepackage{stackengine}
\stackMath
\def\hatgap{2pt}
\def\subdown{-2pt}
\newcommand\reallywidehat[2][]{%
  \renewcommand\stackalignment{l}%
  \stackon[\hatgap]{#2}{%
    \stretchto{%
      \scalerel*[\widthof{$#2$}]{\kern-.6pt\bigwedge\kern-.6pt}%
      {\rule[-5\textheight]{0.1ex}{\textheight}}
    }{0.5ex}
    _{\smash{\belowbaseline[\subdown]{\scriptstyle#1}}}%
  }}

\newtheorem{theorem}{Theorem}

\newtheorem{remark}{Remark}
\newtheorem{definition}{Definition}
\newtheorem{proposition}{Proposition}
\newtheorem{lemma}{Lemma}

\begin{document}
\title{Secure Integrated Sensing and Communication}

\author{Onur G\"unl\"u,~\textit{Member},~\textit{IEEE}, Matthieu~R.~Bloch,~\textit{Senior Member},~\textit{IEEE},
  Rafael~F.~Schaefer,~\textit{Senior Member},~\textit{IEEE}, and Aylin Yener,~\textit{Fellow},~\textit{IEEE}
  \thanks{This work has been supported by the German Federal Ministry of Education and Research (BMBF) under the Grant 16KIS1242, German Research Foundation (DFG) under the Grant SCHA 1944/9-1, U.S. National Science Foundation (NSF) under the grants CCF 1955401 and 2148400, ZENITH Research and Career Development Fund, and the ELLIIT funding endowed by the Swedish government. Parts of the previous versions of these results were presented at the 2022 IEEE International Symposium on Information Theory in Espoo, Finland in \cite{OurISIT2022JCAS}.}
  \thanks{O. G\"unl\"u is with the Information Coding Division, Link{\"o}ping University, 581 83 Link{\"o}ping, Sweden (E-mail: onur.gunlu@liu.se).}
  \thanks{M. Bloch is with the School of Electrical and Computer Engineering, Georgia Institute of Technology, Atlanta, GA 30332 (E-mail: matthieu.bloch@ece.gatech.edu).}
  \thanks{R. F. Schaefer is with the Chair of Information Theory and Machine Learning, Technical University of Dresden, 01062 Dresden, Germany (E-mail: rafael.schaefer@mailbox.tu-dresden.de).}
  \thanks{A. Yener is with the Department of Electrical and Computer Engineering, The Ohio State University, Columbus, OH 43210 (E-mail: yener@ece.osu.edu).}
}

\maketitle

\begin{abstract}
  This work considers the problem of mitigating information leakage between  communication and sensing in systems jointly performing both operations. Specifically, a discrete memoryless state-dependent broadcast channel model is studied in which
  \begin{inparaenum}[(i)]
  \item the presence of feedback enables a transmitter to convey information, while simultaneously performing channel state estimation;
  \item one of the receivers is treated as an eavesdropper whose state should be estimated but which should remain oblivious to part of the transmitted information.
  \end{inparaenum}
  The model abstracts the challenges behind security for joint communication and sensing if one views the channel state as a key attribute, e.g., location. For independent and identically distributed states, perfect output feedback, and when part of the transmitted message should be kept secret, a partial characterization of the secrecy-distortion region is developed. The characterization is exact when the broadcast channel is either physically-degraded or reversely-physically-degraded. The partial characterization is also extended to the situation in which the entire transmitted message should be kept secret. The benefits of a joint approach compared to separation-based secure communication and state-sensing methods are illustrated with binary joint communication and sensing models.
\end{abstract}

%
\begin{IEEEkeywords}
  Secure joint communication and sensing, secure integrated sensing and communication, physical layer security, future communication networks.
\end{IEEEkeywords}

\section{Introduction} \label{sec:intro}
The vision for next generation mobile communication networks includes a seamless integration of the physical and digital world. Key to its success is the network's ability to automatically react to changing environments thanks to tight integration of communication and sensing~\cite{NokiaGuysJCASTutorial}. For instance, a millimeter wave (mmWave) joint communication and radar system can be used to detect a target or to estimate crucial parameters relevant to communication and adapt the communication scheme accordingly~\cite{JCASwithSecurityTutorial}. Integrated sensing and communication (ISAC), also known as joint communication and sensing, techniques are envisioned more broadly as key enablers for a wide range of applications, including connected vehicles and drones \cite{VDEJCASPositionPaper}.

Several information-theoretic studies of ISAC have been initiated, drawing on existing results for joint communication and state estimation~\cite{Zhang_2011,AcademicsJCASTutorial,MassiveMIMOforJCAS}. Motivated by the integration of communication and radar for mmWave vehicular applications,~\cite{MariMichelleGJournalEarlyAccess} considers a model in which messages are encoded and sent through a state-dependent channel with generalized feedback both to reliably communicate with a receiver and to estimate the channel state by using the feedback and transmitted codewords. The optimal trade-off between the communication rate and channel-state estimation distortion is then characterized for memoryless ISAC channels and independent and identically distributed (i.i.d.) channel states that are causally available at the receiver and estimated at the transmitter by using a strictly causal channel output. Follow up works have extended the model to multiple access channels~\cite{MariMACJCAS} and broadcast channels~\cite{MariMichelleGJournalEarlyAccess}.

The nature of ISAC mandates the use of a single modality for the communication and sensing functions so that sensing signals carry information, which then creates situations in which leakage of information may occur. For example, a target illuminated for ranging has the ability to gather potentially sensitive information about the transmitted message~\cite{SecureJCASWireless}. As both sensing and secrecy performance are measured with respect to the signal received at the sensed target, there exists a trade-off between the two~\cite{JCASwithSecurityTutorial,ExtraRef2}. To capture and characterize this trade-off, we extend the ISAC model in \cite{MariMichelleGJournalEarlyAccess} by introducing an eavesdropper in the network. The objective of the transmitter is then to simultaneously communicate reliably with the legitimate receiver, estimate the channel state, and hide a part of the message from the eavesdropper. The channel state is modeled as a two-component state capturing the characteristics of each individual receiver, the feedback is modeled as perfect output feedback for simplicity, and the transmitted message is divided into two parts, only one of which should be kept secret (a setup called partial secrecy in~\cite{RaviZivPartialSecrecyWTC}). The proposed secure ISAC model can be viewed as extensions of the wiretap channel with feedback models~\cite{AhlswedeCaiWTCwithFeedback,AsafCohenWTCwithFeedback,OurJSAITTutorial,HanVinckWTCwithFeedback,he-yener-fbsecrecy,GermanWTCwithGeneralizedFeedback,YHKimWTCwithFeedback,AminGerhardTwoWaySecrecy,Tahmasbi2018}.

\subsection{Summary of Contributions}
Our problem formulation introduces a strong secrecy constraint by considering an eavesdropper whose channel parameters are estimated at the transmitter, but that should be kept ignorant of part of the transmitted message. Even if the state sequence on which the ISAC channel depends is i.i.d., strictly causal channel output feedback improves the secrecy performance. A summary of the main contributions is as follows:

\begin{itemize}
\item We develop inner and outer bounds on the secrecy-distortion region of the secure ISAC model under partial secrecy 
  when i.i.d. channel states are causally available at the corresponding receivers. Our achievability proof leverages the output statistics of random binning (OSRB) method \cite{AhlswedeCsiz,OSRBAmin,RenesRenner}. Our outer bound also holds in the presence of noisy generalized output feedback.
\item We simplify the inner and outer bounds on the secrecy-distortion region when the ISAC channel is physically-degraded or reversely-physically-degraded such that the inner and outer bounds match.
\item We develop inner and outer bounds on the secrecy-distortion region under full secrecy, when the entire transmitted message should be kept secret from the eavesdropper. We characterize the exact secrecy-distortion region under full secrecy when the ISAC channel is physically-degraded or reversely-physically-degraded.
\item We study a binary noiseless ISAC channel example with multiplicative Bernoulli states to illustrate how secure ISAC methods may outperform separation-based secure communication and state-sensing methods. We also consider noisy more-capable ISAC channels to illustrate the effects of noise on an achievable strong secrecy-distortion region.
\end{itemize}

\subsection{Organization}
In Section~\ref{sec:problem_setting}, we introduce the model for secure ISAC under partial secrecy. In Section~\ref{sec:JCASwithPSandPOFResults}, we provide inner and outer bounds on the secrecy-distortion region, specializing them for physically-degraded and reversely-physically-degraded ISAC channels and showing that the bounds match for such channels. In Section~\ref{sec:SingleMess}, we specialize the inner and outer bounds to the full secrecy case. In Section~\ref{sec:JCASexamples}, we illustrate the benefits of integrating security by design into ISAC by evaluating the rate region for a degraded and noiseless ISAC channel with multiplicative Bernoulli states. We also evaluate an achievable region for a noisy ISAC channel with a state-dependent input transmitted through a binary erasure channel (BEC) for the main channel and a binary symmetric channel (BSC) for the eavesdropper's channel, respectively, by establishing the parameter range for which the ISAC channel is more-capable.

\subsection{Notation}
Upper case letters represent random variables and corresponding lower case letters their realizations. A random variable $\displaystyle X$ has probability distribution $\displaystyle P_X$. Calligraphic letters $\displaystyle \mathcal{X}$ denote sets with sizes $\displaystyle |\mathcal{X}|$. A subscript $i$ denotes the position of a variable in a sequence of variables represented by a superscript, e.g., $\displaystyle X^n=X_1,X_2,\ldots, X_i,\ldots, X_n$. $X^{n\setminus i}$ denotes the sequence $X_1,X_2,\ldots,X_{i-1},X_{i+1},\ldots, X_n$ and $X_{i}^k$ denotes $X_{i},X_{i+1},\ldots, X_k$ for integers $i\leq k\leq n$. $[1:J]$ denotes the set $\{1,2,\ldots,J\}$ for an integer $J\geq1$, and $X\sim \text{Unif}[1:J]$ represents a uniform distribution over the set $[1:J]$. $H_b(x)=-x\log x- (1-x)\log (1-x)$ with natural logarithms is the binary entropy function. $X\sim\text{Bern}(p)$ represents a Bernoulli random variable $X$ with probability $p$ of success. $\text{BSC}(\beta)$ denotes a BSC with crossover probability $\beta$ and $\text{BEC}(\gamma)$ a BEC with erasure probability $\gamma$ and erasure symbol $\mathtt{e}$. The $*$ operator denotes the operation $p*\beta = p(1-\beta) + (1-p)\beta$, $\oplus$ is the modulo-$2$ sum, and we define $[a]^+=\max\{a,0\}$ for $a\in\mathbb{R}$.

\section{Problem Definition}\label{sec:problem_setting}
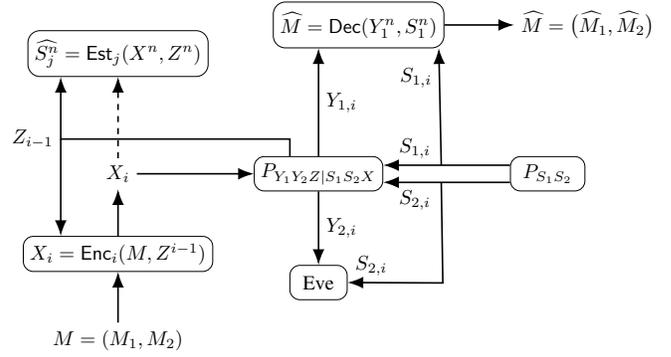
\begin{figure}
  \centering
  \resizebox{\linewidth}{!}{
    \begin{tikzpicture}
      \node (a) at (0,-1.0) [draw,rounded corners = 6pt, minimum width=2.2cm,minimum height=0.8cm, align=left] {$\widehat{S^n_j} = \Est_j(X^n,Z^n)$};
      \node (c) at (3.5,-3.1) [draw,rounded corners = 5pt, minimum width=1.3cm,minimum height=0.6cm, align=left] {$P_{Y_1Y_2Z|S_1S_2X}$};
      \node (state) at (7.5,-3.1) [draw,rounded corners = 5pt, minimum width=1.3cm,minimum height=0.6cm, align=left] {$P_{S_1S_2}$};
      \draw[decoration={markings,mark=at position 1 with {\arrow[scale=1.5]{latex}}},
      postaction={decorate}, thick, shorten >=1.4pt] ($(state.west)+(0,0.15)$) -- ($(c.east)+(0,0.15)$) node [near end, above] {$S_{1,i}$};
      \draw[decoration={markings,mark=at position 1 with {\arrow[scale=1.5]{latex}}},
      postaction={decorate}, thick, shorten >=1.4pt] ($(state.west)+(0,-0.15)$) -- ($(c.east)+(0,-0.15)$) node [near end, below] {$S_{2,i}$};
      \node (b) at (4.2,-0.5) [draw,rounded corners = 6pt, minimum width=2.2cm,minimum height=0.8cm, align=left] {$\widehat{M}=\Dec(Y_1^n,S_1^n)$};
      \node (g) at (3.5,-5) [draw,rounded corners = 5pt, minimum width=1cm,minimum height=0.6cm, align=left] {Eve};
      \draw[decoration={markings,mark=at position 1 with {\arrow[scale=1.5]{latex}}},
      postaction={decorate}, thick, shorten >=1.4pt] ($(state.west)+(-1.2,0.15)$) -- ($(b.south)+(1.4,0)$) node [near end, left] {$S_{1,i}$};
      \draw[decoration={markings,mark=at position 1 with {\arrow[scale=1.5]{latex}}},
      postaction={decorate}, thick, shorten >=1.4pt] ($(state.west)+(-1.2,-0.15)$) -- ($(b.south)+(1.47,-4.1)$) --   ($(g.east)+(0,0.00)$) node [near end, above] {$S_{2,i}$};
      \node (a1) [below of = a, node distance = 2.1cm] {$X_i$};
      \draw[decoration={markings,mark=at position 1 with {\arrow[scale=1.5]{latex}}},
      postaction={decorate}, thick, shorten >=1.4pt] ($(c.north)+(0.0,0)$) -- ($(b.south)-(0.7,0)$) node [midway, right] {$Y_{1,i}$};
      \draw[decoration={markings,mark=at position 1 with {\arrow[scale=1.5]{latex}}},
      postaction={decorate}, thick, shorten >=1.4pt] (a1.east) -- ($(c.west)-(0,0.0)$);
      \draw[decoration={markings,mark=at position 1 with {\arrow[scale=1.5]{latex}}},
      postaction={decorate}, thick, shorten >=1.4pt,dashed] (a1.north) -- ($(a.south)-(0,0.0)$);
      \draw[decoration={markings,mark=at position 1 with {\arrow[scale=1.5]{latex}}},
      postaction={decorate}, thick, shorten >=1.4pt] ($(c.north)-(0.5,0.0)$) -- ($(c.north)-(0.5,-0.3)$) -- ($(c.north)-(4.5,-0.3)$) -- ($(a.south)-(1,0)$);
      \draw[decoration={markings,mark=at position 1 with {\arrow[scale=1.5]{latex}}},
      postaction={decorate}, thick, shorten >=1.4pt] (c.south) -- (g.north) node [midway, right] {$Y_{2,i}$};
      \node (b2) [right of = b, node distance = 4cm] {$\widehat{M}=\big(\widehat{M}_1,\widehat{M}_2\big)$};
      \draw[decoration={markings,mark=at position 1 with {\arrow[scale=1.5]{latex}}},
      postaction={decorate}, thick, shorten >=1.4pt] (b.east) -- (b2.west);
      \node (a2) [below of = a, node distance = 5cm] {$M=(M_1,M_2)$};
      \node (f2) at (0,-4.5) [draw,rounded corners = 5pt, minimum width=1cm,minimum height=0.6cm, align=left] {$X_i=\Enc_i(M,Z^{i-1})$};
      \draw[decoration={markings,mark=at position 1 with {\arrow[scale=1.5]{latex}}},
      postaction={decorate}, thick, shorten >=1.4pt]  (f2.north) -- (a1.south);
      \draw[decoration={markings,mark=at position 1 with {\arrow[scale=1.5]{latex}}},
      postaction={decorate}, thick, shorten >=1.4pt] (a2.north) -- (f2.south) ;
      \draw[decoration={markings,mark=at position 1 with {\arrow[scale=1.5]{latex}}},
      postaction={decorate}, thick, shorten >=1.4pt] ($(c.north)-(4.5,-0.3)$) -- ($(f2.north)-(1,0)$) node [pos=0.0, left] {$Z_{i-1}$};
    \end{tikzpicture}
  }
  \caption{ISAC model under partial secrecy, where only $M_2$ should be kept secret from Eve, for $j=1,2$ and $i~=~[1:n]$. We mainly consider ISAC with perfect output feedback, where $Z_{i-1}=(Y_{1,i-1},Y_{2,i-1})$.}\label{fig:SecureJCASModel}
  \vspace*{-0.4cm}
\end{figure}

We consider the secure ISAC model shown in Fig.~\ref{fig:SecureJCASModel}, which includes a transmitter equipped with a state estimator, a legitimate receiver, and an eavesdropper (Eve). The transmitter attempts to reliably transmit a uniformly distributed message $M=(M_1,M_2)\in \mathcal{M}=\mathcal{M}_1\times \mathcal{M}_2$ through a memoryless state-dependent ISAC channel with known statistics $P_{Y_1Y_2Z|S_1S_2X}$ and i.i.d. state sequence $(S_1^n,S_2^n)\in\mathcal{S}_1^n\times\mathcal{S}_2^n$ generated according to a known joint probability distribution $P_{S_1S_2}$. The transmitter calculates the channel inputs $X^n$ as $X_i=\Enc_i(M,Z^{i-1})\in \mathcal{X}$ for all $i=[1:n]$, where $\Enc_i(\cdot)$ is an encoding function and $Z^{i-1}\in\mathcal{Z}^{i-1}$ is the delayed channel output feedback. The legitimate receiver that observes $Y_{1,i}\in\mathcal{Y}_1$ and $S_{1,i}\in\mathcal{S}_1$ for all channel uses $i=[1:n]$ should reliably decode both $M_1$ and $M_2$ by forming the estimate $\widehat{M}=\Dec(Y_1^n,S_1^n)$, where $\Dec(\cdot)$ is a decoding function. The eavesdropper that observes $Y_{2,i}\in\mathcal{Y}_2$ and $S_{2,i}\in\mathcal{S}_2$ should be kept ignorant of $M_2$. Finally, the transmitter estimates the state sequence $(S_1^n,S_2^n)$ as $\widehat{S^n_j} = \Est_j(X^n,Z^n)\in\reallywidehat[n]{\mathcal{S}_j}$ for $j=1,2$, where $\Est_j(\cdot,\cdot)$ is an estimation function. All sets $\mathcal{S}_1$, $\mathcal{S}_2$, $\widehat{\mathcal{S}}_1$, $\widehat{\mathcal{S}}_2$, $\mathcal{X}$, $\mathcal{Y}_1$, $\mathcal{Y}_2$, and $\mathcal{Z}$ are finite.

This channel model can be viewed as an abstraction of ISAC with a multi-functional phased array, in which a transmitter exploits backscattered waveforms (the channel output feedback $Z^{i-1}$) to infer information about the states ($S_{1,i}$ and $S_{2,i}$) that affect the transmission in the directions of a legitimate receiver and an eavesdropper.

For simplicity, we consider the perfect output feedback case, in which for all $i=[2:n]$ we have
\begin{align}
  Z_{i-1}=(Y_{1,i-1},Y_{2,i-1}).\label{eq:POFcondition}
\end{align}
Although the perfect output feedback is explicitly used in our achievability proofs, some of our converse results hold for generalized feedback. Furthermore, the fundamental insights gained from our results can be used to tackle generalized feedback scenarios, for which identifying closed-form characterizations becomes challenging; see, e.g., \cite{GermanWTCwithGeneralizedFeedback}. We next define the strong secrecy-distortion region for the problem of interest.

\begin{definition}\label{def:systemmodel}
  \normalfont A secrecy-distortion tuple $(R_{1}, R_{2},D_{1},D_{2})$ is \emph{achievable} under partial secrecy if, for any $\delta\!>\!0$, there exist $n\!\geq\!1$, one encoder, one decoder, and two estimators $\Est_j(X^n,Y_1^n,Y_2^n )=\widehat{S_j^n}$, $j\in\{1,2\}$, such that
  \begin{align}
    & \frac{1}{n}\log |\mathcal{M}_j|\geq R_j -\delta\quad\;\;\;\;\;\text{for } j\!=\!1,2\;\;&&\!\!\!\!\!(\text{rates})\label{eq:rates_cons}\\ 
    &\Pr\big[(M_1,M_2) \neq (\widehat{M}_1,\widehat{M}_2)\big] \leq \delta&&\!\!\!\!\! (\text{reliability})\label{eq:reliability_cons}\\
    &I(M_2;Y^n_2|S_2^n) \leq \delta&&\!\!\!\!\!(\text{strong secrecy})\label{eq:secrecyleakage_cons}\\
    &\mathbb{E}\big[d_j(S_j^n,\widehat{S_j^n})\big] \!\leq\! D_j\!+\!\delta\;\;\;\;\;\;\text{for } j\!=\!1,2\;\;&&\!\!\!\!\!(\text{distortions})\label{eq:distortion_consts}
  \end{align}
  where $d_j(s^n,\widehat{s^n})=\frac{1}{n}\sum_{i=1}^nd_j(s_i,\widehat{s}_i)$ for $j\!=\!1,2$ are bounded per-letter distortion metrics.
  
  The secrecy-distortion region $\mathcal{R}_{\textnormal{PS,POF}}$ is the closure of the set of all achievable tuples under partial secrecy and perfect output feedback. \hfill $\lozenge$
\end{definition}

The use of per-letter distortion metrics $d_j(\cdot,\cdot)$ in conjunction with i.i.d. states reduces the problem to the characterization of a rate distortion region~\cite{MariMichelleGJournalEarlyAccess,MariMACJCAS}; in fact, past observations are independent of present and future ones, lending the transmitter no state prediction ability to adapt its transmission on the fly. Analyzing ISAC models with memory leads to conceptually different results; see, e.g.,~\cite{Chang2022Rate,WillemsJCAS,Wu2022Joint}. In practical ISAC applications, only a part of the channel parameters might be relevant for the transmitter \cite{ISACMACChinesegroup}. Our results can be extended for such cases by adapting the estimator functions used and not requiring an estimation of the exact state.

\begin{remark}\label{rem:secrecyconstraintwithoutcond}
  \normalfont The strong secrecy condition (\ref{eq:secrecyleakage_cons}) is equivalent to $I(M_2;Y^n_2,S_2^n) \leq \delta$ since the transmitted message is independent of the state sequence and $I(M_2;Y^n_2,S_2^n)=I(M_2;Y^n_2|S_2^n)$.
\end{remark}

\section{ISAC Under Partial secrecy}\label{sec:JCASwithPSandPOFResults}
We next present inner and outer bounds on the secrecy-distortion region $\mathcal{R}_{\textnormal{PS,POF}}$. 

\begin{proposition}[Inner Bound]\label{prop:InnerforPSPOF}
  The region $\mathcal{R}_{\textnormal{PS,POF}}$ includes the union over all joint distributions $P_{UVX}$ of the rate tuples $(R_{1}, R_{2},D_1,D_2)$ such that
  \begin{align}
    &R_{1}\leq I(U;Y_1|S_1)\label{eq:achR1}\\
    &R_{2}\leq \min\{R_{2}^{\prime}, (I(V;Y_1|S_1)-R_1)\}\label{eq:achR2}\\
    & D_j\geq \mathbb{E}[d_j(S_j,\widehat{S}_j))]\qquad\qquad  \text{for }j=1,2\label{eq:achdistortion1and2}
  \end{align}
  where
  \begin{align}
    &P_{UVXY_1Y_2S_1S_2} = P_{U|V}P_{V|X}P_XP_{S_1S_2}P_{Y_1Y_2|S_1S_2X}\label{eq:jointprobiid},\\
    &R_{2}^{\prime}=[I(V;Y_1|S_1,U)-I(V;Y_2|S_2,U)]^{+}\nonumber\\
    &\qquad\qquad+H(Y_1|Y_2,S_2,V)\label{eq:R2primedef}
  \end{align}
  and one can apply the per-letter estimators $\Est_j(x,y_1,y_2)=~\hat{s}_j$ for $j=1,2$ such that 
  \begin{align}
    &\Est_j(x,y_1,y_2)\nonumber\\
    &=\mathop{\textnormal{argmin}}_{\tilde{s}\in\widehat{\mathcal{S}}_j} \sum_{s_j\in\mathcal{S}_j}P_{S_j|XY_1Y_2}(s_j|x,y_1,y_2)\; d_j(s_j,\tilde{s}).\label{eq:deterministicest} 
  \end{align}
  One can limit $|\mathcal{U}|$ to 
  \begin{align}
    \min\{|\mathcal{X}|,\;|\mathcal{Y}_1|\!\cdot\!|\mathcal{S}_1|,\;|\mathcal{Y}_2|\!\cdot\!|\mathcal{S}_2|\}\!+\!2\label{eq:cardUforPS}
  \end{align}
  and $|\mathcal{V}|$ to
  \begin{align}
    &(\min\{|\mathcal{X}|,\;|\mathcal{Y}_1|\!\cdot\!|\mathcal{S}_1|,\;|\mathcal{Y}_2|\!\cdot\!|\mathcal{S}_2|\}\!+\!2)\nonumber\\
    &\qquad\cdot(\min\{|\mathcal{X}|,\;|\mathcal{Y}_1|\!\cdot\!|\mathcal{S}_1|,\;|\mathcal{Y}_2|\!\cdot\!|\mathcal{S}_2|\}\!+\!1).\label{eq:cardVforPS}
  \end{align}
\end{proposition}

Proposition~\ref{prop:InnerforPSPOF} can be interpreted as follows. The rate $R_1$ in (\ref{eq:achR1}) represents a rate of a public message that could be decoded by the eavesdropper. The rate $R_2^{\prime}$ in (\ref{eq:R2primedef}) represents the rate of a secret message superposed to the public message. $R_2^{\prime}$ is itself the sum of two terms: a first term representing a wiretap-coding rate against an eavesdropper observing the public message; a second term representing a secret key rate extracted from the feedback channel and used as a one-time pad. The operator $[\cdot]^+$ in (\ref{eq:R2primedef}) indicates that the decoder should have an advantage over the eavesdropper to apply wiretap-coding methods. The minimum operator in (\ref{eq:achR2}) merely indicates that the secrecy rate cannot exceed the reliable communication rate. Most importantly, Proposition~\ref{prop:InnerforPSPOF} suggests that secure ISAC systems benefit from the inherent presence of the feedback link, which allows the transmitter to develop situational awareness and extract secret keys from the wireless environment.

\begin{IEEEproof}[Proof of Proposition~\ref{prop:InnerforPSPOF}]
  We use the OSRB method \cite{OSRBAmin,RenesRenner} for the achievability proof, applying the steps in~\cite[Section~1.6]{BlochLectureNotes2018}; see also \cite{YenerOSRB}. Following~\cite{OSRBAmin}, we shall first define an operationally dual source coding problem to the original ISAC problem, along with a coding scheme called \emph{Protocol~A}, for which reliability and secrecy analyses are conducted. These analyses consist of imposing bounds on the sizes of the bins assigned to $n$-letter sequences such that either a sequence reconstruction constraint is satisfied via \cite[Lemma 1]{OSRBAmin} by using a Slepian-Wolf~\cite{SW} decoder or mutual independence and uniformity constraints are satisfied via \cite[Theorem 1]{OSRBAmin} by using privacy amplification. We shall next define a randomized coding scheme, called \emph{Protocol~B}, for the original ISAC problem and show that the joint probability distributions induced by Protocols~A and B are almost equal, allowing us to invert the source code proposed for Protocol~A to construct a channel code for Protocol~B. The achievability proof shall finally follow by derandomizing Protocol B and chaining multiple uses of Protocol B over several blocks such that chaining does not affect the secrecy and reliability performance. 
  
  \textbf{Protocol~A} (dual source coding problem): We consider a secret key agreement model for an i.i.d. source with distribution $P_{UVXY_1Y_2S_1S_2}$ as in~(\ref{eq:jointprobiid}), in which a source encoder observing $(U^n,V^n,X^n)\in\mathcal{U}^n\times\mathcal{V}^n\times\mathcal{X}^n$ assigns random bin indices $M\in\mathcal{M}=\mathcal{M}_1\times\mathcal{M}_2$ and $F\in\mathcal{F}$ to its observations. The index pair $M=(M_1,M_2)$ should be reliably reconstructed at a legitimate source receiver observing $(Y_1^n, S_1^n)\in\mathcal{Y}_1^n\times \mathcal{S}_1^n$ and $F$ to satisfy (\ref{eq:reliability_cons}), while keeping $M_2$ secret from an eavesdropper observing $(Y_2^n, S_2^n)\in\mathcal{Y}_2^n\times \mathcal{S}_2^n$ and $F$ to satisfy the strong secrecy constraint (\ref{eq:secrecyleakage_cons}). Furthermore, we assume that $P_{UV|X}$ has been chosen so that distortion constraints (\ref{eq:distortion_consts}) can be satisfied, i.e., there exist associated per-letter estimators $\Est_j(x,y_1,y_2)=\widehat{S_j}$ for $j=1,2$ such that
  \begin{align}
    &\mathbb{E}[d_j(S_j^n,\Est_j^n(X^n,Y^n_1,Y_2^n))]\leq D_j+\epsilon_n^{\prime}\label{eq:assumedperletterestimatorach}
  \end{align}
  where $\epsilon_n^{\prime}>0$ such that $\epsilon_n^{\prime}\rightarrow 0$ when $n\rightarrow\infty$.

  Formally, we construct Protocol A as follows. To each sequence $u^n$, we independently and uniformly assign two random bin indices $(F_{\text{u}},W_{\text{u}})$ such that 	$F_{\text{u}}\in[1:2^{n\widetilde{R}_{\text{u}}}]$ and $W_{\text{u}}\in[1:2^{nR_{\text{u}}}]$. Furthermore, to each sequence $v^n$, we independently and uniformly assign three random indices $(F_{\text{v}},W_{\text{v}},L_{\text{v}})$ such that $F_{\text{v}}\in[1:2^{n\widetilde{R}_{\text{v}}}]$, $W_{\text{v}}\in[1:2^{nR_{\text{v}}}]$, and $L_{\text{v}}\in[1:2^{n\overline{R}_{\text{v}}}]$. Finally, to each sequence $y_1^{n}$, we independently and uniformly assign a random index $L_{\text{y}_1}\in[1:2^{n\overline{R}_{\text{y}_1}}]$ with $\overline{R}_{\text{y}_1} = \overline{R}_{\text{v}}$. The index pair $F=(F_{\text{u}}, F_{\text{v}})$ shall be transmitted publicly to allow the reliable reconstruction of the source encoder observations at the legitimate source receiver. The index tuple $W=(W_{\text{u}}, W_{\text{v}},L_{\text{v}})$ represent indices that can then be reliably computed at the source receiver, and we shall impose a secrecy constraint on $W_{\text{v}}$. The index $L_{\text{y}_1}$, which is derived from $y_1^n$ and therefore known at both the source encoder and the legitimate receiver, shall also be subject to a secrecy constraint. We finally set
  \begin{align}
    M_1 = W_{\text{u}}\qquad \text{ and }\qquad  M_2= (W_{\text{v}}, L_{\text{v}}\oplus L_{\text{y}_1})\label{eq:assignmentofM1M2ach}.
  \end{align}
  We next develop conditions on the bin sizes to ensure the required reliability and secrecy constraints. Using a SW decoder, the expected value (over the random bin assignments) of the probability of incorrectly reconstructing $U^n$ from $(Y_1^n,S_1^n,F_{\text{u}})$  vanishes exponentially fast when $n\rightarrow~\infty$ if we have \cite[Lemma 1]{OSRBAmin}
  \begin{align}
    \widetilde{R}_{\text{u}} > H(U|Y_1,S_1).\label{eq:reconstrU}
  \end{align}	
  Similarly, the probability of incorrectly reconstructing $V^n$ from $(Y_1^n,S_1^n,F_{\text{v}}, U^n)$ vanishes exponentially fast if
  \begin{align}
    \widetilde{R}_{\text{v}} > H(V|Y_1,S_1,U).\label{eq:reconstrV}
  \end{align}
  Using privacy amplification~\cite[Theorem 1]{OSRBAmin}, the expected value of the variational distance between the joint probability distributions $\text{Unif}[1\!\!:\!2^{nR_{\text{u}}}]\cdot \text{Unif}[1\!\!:\!2^{n\widetilde{R}_{\text{u}}}]$ and $P_{W_{\text{u}}F_{\text{u}}}$ vanishes exponentially fast when $n\rightarrow\infty$ if
  \begin{align}
    R_{\text{u}}+\widetilde{R}_{\text{u}}<H(U).\label{eq:independenceofFuWu}
  \end{align}
  With a slight abuse of terminology, we shall concisely say that the indices $F_{\text{u}}$ and $W_{\text{u}}$ then become almost independent and uniformly distributed. Similarly, the indices $F_{\text{v}}$ and $W_{\text{v}}$ become almost independent of $(Y_2^n, S_2^n, U^n)$ and uniformly distributed if
  \begin{align}
    R_{\text{v}}+\widetilde{R}_{\text{v}}<H(V|Y_2,S_2,U)\label{eq:independenceofFvWv}
  \end{align} 
  and the index $L_{\text{y}_1}$ becomes almost independent of $\big(Y_2^n, S_2^n, V^n, U^n\big)$ and uniformly distributed if
  \begin{align}
    \overline{R}_{\text{y}_1}&= \overline{R}_{\text{v}} < H(Y_1|Y_2,S_2,V,U)\overset{(a)}{=}H(Y_1|Y_2,S_2,V)\label{eq:independenceofYindexnewwwwwwww}
  \end{align} 
  where $(a)$ follows because $U-V-(Y_1,Y_2,S_2)$ form a Markov chain. Note that $L_\text{v}\oplus L_{\text{y}_1}$ is then also almost independent of $\big(Y_2^n, S_2^n,$ $ V^n, U^n\big)$ and uniformly distributed.

  Finally, $(F_{\text{u}},W_{\text{u}},F_{\text{v}},W_{\text{v}},L_{\text{v}})$ are almost mutually independent and uniformly distributed if 
  \begin{align}
    R_{\text{u}}+\widetilde{R}_{\text{u}}+R_{\text{v}}+\widetilde{R}_{\text{v}} + \overline{R}_{\text{v}}< H(U,V).\label{eq:sumindependence}
  \end{align}

  Assuming $(I(V;Y_1,S_1|U)-I(V;Y_2,S_2|U))>0$ and $\epsilon>~0$ small enough, a specific choice of rates that satisfies all conditions above is
  \begin{align}
    &\widetilde{R}_{\text{u}} = H(U|Y_1,S_1)+\epsilon\\
    &R_{\text{u}} = H(U) - H(U|Y_1,S_1) -2\epsilon = I(U;Y_1|S_1)-2\epsilon\\
    &\widetilde{R}_{\text{v}}= H(V|Y_1,S_1,U)+\epsilon\\
    &R_{\text{v}} = H(V|Y_2,S_2,U) - H(V|Y_1,S_1,U) - 2\epsilon \nonumber\\
    &\;\;\;\;\,= I(V;Y_1|S_1,U)-I(V;Y_2|S_2,U)\!-\!2\epsilon\\
    &\overline{R}_{\text{v}} = \min\{(H(Y_1|Y_2,S_2,U,V) -\epsilon), \nonumber\\
    &\qquad\qquad\qquad(H(U,V)-H(U)-H(V|Y_2,S_2,U))\}\\
    &\;\;\;\;\,= \min \{(H(Y_1|Y_2,S_2,V) -\epsilon), I(V;Y_2|S_2,U)\}
  \end{align}
  where we have repeatedly used the independence of $(U,V)$ with $(S_1,S_2)$ to simplify the expressions. If $(I(V;Y_1,S_1|U)-~I(V;Y_2,S_2|U))\!\leq\! 0$, one should set
  \begin{align}
    &R_{\text{v}}=0\nonumber\\
    &\overline{R}_{\text{v}}= \min\{H(Y_1|Y_2,S_2,V), I(V;Y_1|S_1,U)\}-\epsilon.
  \end{align}
  Combining the above of choices with our definition $R_1=~R_{\text{u}}$ and $R_2=R_{\text{v}}+\overline{R}_{\text{v}}$, we recover the rate conditions in Proposition~\ref{prop:InnerforPSPOF}, satisfying the reliability condition~(\ref{eq:reliability_cons}) and the secrecy condition~(\ref{eq:secrecyleakage_cons}). Finally, we  consider the distortion constraints (\ref{eq:distortion_consts}) on the channel state estimations. All $(u^n,v^n,x^n,y_1^n,y_2^n,s_1^n,s_2^n)$ tuples are in the jointly typical set with high probability and, by applying the law of total expectation to bounded distortion metrics and from the typical average lemma \cite[pp.~26]{Elgamalbook}, distortion constraints (\ref{eq:distortion_consts}) are satisfied. Furthermore, without loss of generality one can use the deterministic per-letter estimators in (\ref{eq:deterministicest}) and the proof follows from the proof of \cite[Lemma~1]{MariMichelleGJournalEarlyAccess} by replacing $(S,Z,\hat{S},d)$ with $(S_j,(Y_1,Y_2),\widehat{S}_j,d_j)$, respectively, since $\widehat{S_j}-(X,Y_1,Y_2)-S_j$ form a Markov chain for all $j=1,2$.

  This concludes the construction and analysis of Protocol~A. Note that Protocol~A induces a joint probability distribution $P_{M_1M_2FX^n}P_{S_1^nS_2^n}P_{Y_1^nY_2^n|S_1^nS_2^nX^n}$ that is asymptotically indistinguishable in variational distance from a distribution of the form
  \begin{align}
    &\text{Unif}[1:2^{n(\widetilde{R}_{\text{u}}+\widetilde{R}_{\text{v}})}]\cdot\text{Unif}[1:2^{n(R_{\text{u}}+R_{\text{v}}+\overline{R}_{\text{v}})}]\nonumber\\
    &\qquad\cdot P_{S_1^nS_2^n}P_{X^n|M_1M_2F}P_{Y_1^nY_2^n|S_1^nS_2^nX^n}.\label{eq:FandM1M2uniformProbDistr}
  \end{align}

  \textbf{Protocol~B} (random channel coding for the original problem): We now transform Protocol~A into another protocol that is suited to the original ISAC problem.
  Assume that the index pair $F = (F_{\text{u}},F_{\text{v}})$ is generated uniformly at random and disclosed to all parties ahead of the transmission and that the transmitter and the receiver share a secret key $K\in~ \text{Unif}[1:~2^{\overline{R}_{\text{y}_1}}]$. Then, the transmitter encodes a uniformly distributed message $M_1 = W_{\text{u}}$ and another uniformly distributed message $M_2=(W_{\text{v}},L_{\text{v}}\oplus K)$ according to the distribution $P_{X^n|M_1M_2F}$, defined by Protocol~A. Note that Protocol~B induces a joint probability distribution as in (\ref{eq:FandM1M2uniformProbDistr}), which is, as argued above, asymptotically indistinguishable in variational distance from the distribution $P_{M_1M_2FX^n}P_{S_1^nS_2^n}P_{Y_1^nY_2^n|S_1^nS_2^nX^n}$, induced by Protocol~A. In other words, Protocol~B guarantees the exact same asymptotic performance as Protocol~A in terms of secrecy, reliability, and distortions subject to the same rate constraints.

  \textbf{Derandomizing and Chaining Protocol~B}. To conclude the achievability, two aspects of Protocol B remain to be fixed to obtain a code for the original ISAC model:
  \begin{inparaenum}[i)]
  \item the public transmission of the index $F$ should be removed; and
  \item the use of the secret key $K$ should be removed.
  \end{inparaenum}
  Following \cite{OSRBAmin}, one can argue that there exists a fixed index $F=f$ such that Protocol~B retains its properties, eliminating the need for a public discussion. We skip this standard step for brevity. Most importantly, note that Protocol A allows us to generate a secret key $L_{\text{y}_1}$. Consequently, if one were to use Protocol~B in a block Markov fashion chained over multiple blocks, the key generated in block $b\geq 1$ can be used as the key for one-time padding in block $(b+1)$, removing the need for a secret key in Protocol~$B$.
  
  Formally, assume that we repeat Protocol~B over $B$ blocks indexed by $b\in[1:B]$. In every block $b$, we denote the messages by a superscript $b$. In particular, the transmitter attempts to transmit messages $W_{\text{u}}^b$, $W_{\text{v}}^b$, and $L_{\text{v}}^b$, as well as generate a key $L_{\text{y}_1}^{b}$. In this section alone, we also denote a sequence of variables across blocks $k$ through $\ell$ ($\ell\geq k$) by the superscript $k:\ell$, e.g., $M_1^{k:\ell}$. In the first block, no message is transmitted and only a key $L_{\text{y}_1}^1$ is generated. In every subsequent block $b\in[2:B]$, the encoder uses Protocol~B to transmit a public message $M_1^b=W_{\text{u}}^b$ and a secret message $M_2^b = (W_{\text{v}}^b,L_{\text{v}}^b\oplus L_{\text{y}_1}^{b-1})$, and generates a key $L_{\text{y}_1}^b$. A union bound shows that the asymptotic reliability performance 
  \begin{align}
    \lim_{n\to\infty} \Pr\left[\Big\{\widehat{M}_1^{1:B}\neq M_1^{1:B}\text{ or } \widehat{M}_2^{1:B}\neq M_2^{1:B}\Big\}\right]
  \end{align}
  is not affected by the chaining. The proof that secrecy is not affected by the chaining requires a bit more care, as we need to show that $I(W_{\text{v}}^{1:B},L_{\text{v}}^{1:B};\mathbf{Y}_2^{1:B},\mathbf{S}_2^{1:B})$, where the bold-face letters represent $n$-letter random variables, vanishes across all blocks. This can be done by adapting the approach of \cite{Sasoglu2013} and \cite{Chou2014d} as we show next. To simplify notation, we set $W^b=W_{\text{v}}^{b}$, $L^{b}=L_{\text{v}}^b$, $K^b =L_{\text{y}_1}^b$, and $\mathbf{Z}^{b}=(\mathbf{Y}_2^{b},\mathbf{S}_2^{b})$.

  We have
  \begin{align*}
    &I(W^{1:B},L^{1:B};\mathbf{Z}^{B})\nonumber\\
    & =  \sum_{b=1}^{B-1}\left( I(W^{1:B},L^{1:B};\mathbf{Z}^{1:b+1}) -  I((W^{1:B},L^{1:B};\mathbf{Z}^{1:b})\right) \nonumber\\
    &\qquad + I(W^{1:B},L^{1:B};\mathbf{Z}^{1})\nonumber\\
    &\overset{(a)}{=}  \sum_{b=1}^{B-1}\left( I(W^{1:B},L^{1:B};\mathbf{Z}^{1:b+1}) -  I((W^{1:B},L^{1:B};\mathbf{Z}^{1:b})\right)
  \end{align*}
  where $(a)$ follows since $I(W^{1:B},L^{1:B};\mathbf{Z}^{1}) = I(W^{1},L^{1};\mathbf{Z}^{1})=0$ by definition because no message is transmitted in the first block. Focusing on every term in the sum for a given index $b$, we obtain
  \begin{align}
    &(I(W^{1:B},L^{1:B};\mathbf{Z}^{1:b+1}) -  I(W^{1:B},L^{1:B};\mathbf{Z}^{1:b}))\nonumber\\
    &\phantom{=}= I(W^{1:B},L^{1:B};\mathbf{Z}^{b+1}|\mathbf{Z}^{1:b})\nonumber\\
    &\phantom{=}= I(W^{1:b+1},L^{1:b+1};\mathbf{Z}^{b+1}|\mathbf{Z}^{1:b})\nonumber\\
    &\qquad\qquad +I(W^{b+2:B},L^{b+2:B};\mathbf{Z}^{b+1}|\mathbf{Z}^{1:b},W^{1:b+1},L^{1:b+1})\nonumber\\
    &\phantom{=}\leq I(W^{1:b+1},L^{1:b+1},\mathbf{Z}^{1:b};\mathbf{Z}^{b+1})\nonumber\\
    &\qquad\qquad +I(W^{b+2:B},L^{b+2:B};\mathbf{Z}^{1:b+1},W^{1:b+1},L^{1:b+1})\nonumber\\
    &\phantom{=}\stackrel{(a)}{=} I(W^{1:b+1},L^{1:b+1},\mathbf{Z}^{1:b};\mathbf{Z}^{b+1})\nonumber\\
    &\phantom{=}=I(W^{b+1},L^{b+1};\mathbf{Z}^{b+1})\nonumber\\
    &\qquad\qquad  +I(W^{1:b},L^{1:b},\mathbf{Z}^{1:b};\mathbf{Z}^{b+1}|W^{b+1},L^{b+1})\nonumber\\
    &\phantom{=}\stackrel{(b)}{=}I(W^{b+1},L^{b+1};\mathbf{Z}^{b+1})\nonumber\\
    &\qquad\qquad +I(W^{1:b},L^{1:b},\mathbf{Z}^{1:b};\mathbf{Z}^{b+1},W^{b+1},L^{b+1})\nonumber\\
    &\phantom{=}\stackrel{(c)}{\leq}I(W^{b+1},L^{b+1};\mathbf{Z}^{b+1})\nonumber\\
    &\qquad\qquad +I(W^{1:b},L^{1:b},\mathbf{Z}^{1:b};\mathbf{Z}^{b+1},W^{b+1},L^{b+1},K^b)\nonumber\\
    &\phantom{=}=I(W^{b+1},L^{b+1};\mathbf{Z}^{b+1})+I(W^{1:b},L^{1:b},\mathbf{Z}^{1:b};K^b)\nonumber\\
    &\qquad\qquad +I(W^{1:b},L^{1:b},\mathbf{Z}^{1:b};\mathbf{Z}^{b+1},W^{b+1},L^{b+1}|K^b)\nonumber\\
    &\phantom{=}\stackrel{(d)}{=}I(W^{b+1},L^{b+1};\mathbf{Z}^{b+1})+I(W^{1:b},L^{1:b},\mathbf{Z}^{1:b};K^b)\nonumber\\
    &\phantom{=}\stackrel{(e)}{\leq}I(W^{b+1},L^{b+1};\mathbf{Z}^{b+1})+I(W^{1:b},L^{1:b},\mathbf{Z}^{1:b},K^{b-1};K^b)\nonumber\\
    &\phantom{=}\stackrel{(f)}{=}\!I(W^{b+1},L^{b+1};\mathbf{Z}^{b+1})\!+\!I(W^{b},L^{b},\mathbf{Z}^{b},K^{b-1};K^b)
  \end{align}
  where $(a)$ follows because the future messages $(W^{b+2:B},L^{b+2:B})$ are independent of past messages and observations $(\mathbf{Z}^{1:b+1},W^{1:b+1},L^{1:b+1})$, $(b)$ follows similarly because future messages $(W^{b+1},L^{b+1})$ are independent of past messages and observations $(W^{1:b},L^{1:b},\mathbf{Z}^{1:b})$, $(c)$ follows by introducing the key $K^b$ generated in block $b$ as an attempt to break dependence across blocks, $(d)$ follows because $I(W^{1:b},L^{1:b},\mathbf{Z}^{1:b};\mathbf{Z}^{b+1},W^{b+1},L^{b+1}|K^b)=0$ since
  \begin{align}
      (W^{1:b},L^{1:b},\mathbf{Z}^{1:b}) - K_b - (\mathbf{Z}^{b+1},W^{b+1},L^{b+1})
  \end{align}
  form a Markov chain, $(e)$ follows by introducing $K^{b-1}$ in an effort to break again dependence, and $(f)$ follows since $I(W^{1:b-1},L^{1:b-1},\mathbf{Z}^{1:b-1};K^b|W^{b},L^{b},\mathbf{Z}^{b}K^{b-1})=0$ since
  \begin{align}
      (W^{1:b-1},L^{1:b-1},\mathbf{Z}^{1:b-1})-(W^{b},L^{b},\mathbf{Z}^{b},K^{b-1})-K_b
  \end{align}
  form a Markov chain. Consequently, we obtain
  \begin{align*}
    &I(W^{1:B},L^{1:B};\mathbf{Z}^{B})\nonumber\\
    & \leq \sum_{b=1}^{B-1}\left(I(W^{b+1},L^{b+1};\mathbf{Z}^{b+1})+I(W^{b},L^{b},\mathbf{Z}^{b},K^{b-1};K^b)\right).
  \end{align*}
  All that remains to confirm is that in each block, $I(W^{b+1},L^{b+1};\mathbf{Z}^{b+1})$ and $I(W^{b},L^{b},\mathbf{Z}^{b},K^{b-1};K^b)$ are asymptotically vanishing. For the first term, note that
  \begin{align}
    &I(W^{b+1},L^{b+1};\mathbf{Z}^{b+1}) \nonumber\\
    &= I(W^{b+1};\mathbf{Z}^{b+1}) + I(L^{b+1};\mathbf{Z}^{b+1}|W^{b+1})\nonumber\\
    &\stackrel{(a)}{=}I(W^{b+1};\mathbf{Z}^{b+1}) + I(L^{b+1};\mathbf{Z}^{b+1},W^{b+1})\nonumber\\
    &\stackrel{(b)}{\leq} I(W^{b+1};\mathbf{Z}^{b+1}) + I(L^{b+1};L^{b+1}\oplus K^{b})\nonumber
            \end{align}
\begin{align}
    &\stackrel{(c)}{=} I(W^{b+1};\mathbf{Z}^{b+1}) + n \overline{R}_\text{v} - H(K^b)
  \end{align}
  where $(a)$ follows by independence of $W^{b+1}$ and $L^{b+1}$, $(b)$ follows because of the Markov chain $L^{b+1}-(L^{b+1}\oplus K^{b})-(\mathbf{Z}^{b+1},W^{b+1})$, and $(c)$ follows because $I(L^{b+1};L^{b+1}\oplus K^{b}) = (H(L^{b+1}\oplus K^{b}) - H(L^{b+1}\oplus K^{b}|L^{b+1})) \leq (n \overline{R}_\text{v} - H(K^b))$. Looking back at Protocols A and B, the secrecy of $W^{b+1}$ and the uniformity of $K^b$ guarantee that $I(W^{b+1};\mathbf{Z}^{b+1})$ and $(n \overline{R}_\text{v} - H(K^b))$ are exponentially vanishing as $n\to\infty$; see~(\ref{eq:independenceofFvWv}) and~(\ref{eq:independenceofYindexnewwwwwwww}). Similarly, for the second term, the secrecy of $K^b$ guarantees directly that $I(W^{b},L^{b},\mathbf{Z}^{b},K^{b-1};K^b)$ is exponentially vanishing as $n\to\infty$; see~(\ref{eq:independenceofYindexnewwwwwwww}).

  Lastly, one can check that the chaining over $B$ blocks has a negligible effect on the rate of the coding scheme.
\end{IEEEproof}

\begin{proposition}[Outer Bound]\label{prop:OuterforPSPOF}
  The region $\mathcal{R}_{\textnormal{PS,POF}}$ is included in the union over all joint distributions $P_{UVX}$ of the rate tuples $(R_1,R_2,D_1,D_2)$ satisfying 
  \begin{align}
    & D_j\geq \mathbb{E}[d_j(S_j,\widehat{S}_j))]\qquad\qquad  \text{for }j=1,2\label{eq:achdistortion1and2_new}\\
    &R_{1}\leq I(V;Y_1|S_1)\label{eq:achR1degradedbutnowouterProp2}\\
    &R_2\leq \min\Big\{\big(H(Y_1,S_1|Y_2,S_2) \!-\! H(S_1|Y_1,Y_2,S_2,V)\big),\nonumber\\
    &\qquad\qquad\qquad \big(I(V;Y_1|S_1)-R_1\big)\Big\}    \label{eq:convR2}
  \end{align} 
  where we have 
  \begin{align}
  		&P_{UVXY_1Y_2S_1S_2} = P_{U|V}P_{V|X}P_XP_{S_1S_2}P_{Y_1Y_2|S_1S_2X}\label{eq:jointprobiid_new},\\
  		 &\Est_j(x,y_1,y_2)\nonumber\\
  		 &\;=\mathop{\textnormal{argmin}}_{\tilde{s}\in\widehat{\mathcal{S}}_j} \sum_{s_j\in\mathcal{S}_j}P_{S_j|XY_1Y_2}(s_j|x,y_1,y_2)\; d_j(s_j,\tilde{s}).\label{eq:deterministicest_new} 
  \end{align}
 One can limit $|\mathcal{V}|$ to 
  \begin{align}
    \min\{|\mathcal{X}|,\;|\mathcal{Y}_1|\!\cdot\!|\mathcal{S}_1|,\;|\mathcal{Y}_2|\!\cdot\!|\mathcal{S}_2|\}\!+\!1\label{eq:cardVplus1}.
  \end{align}
\end{proposition}

\begin{remark}
  \normalfont Since we consider perfect feedback as in~(\ref{eq:POFcondition}), the outer bound proposed in Proposition~\ref{prop:OuterforPSPOF} is also valid for the general ISAC problem depicted in Fig.~\ref{fig:SecureJCASModel}, in which the feedback $Z_{i-1}$ can be a noisy version of $(Y_{1,i-1},Y_{2,i-1})$.
\end{remark}

\begin{IEEEproof}[Proof of Proposition~\ref{prop:OuterforPSPOF}]
  Assume that for some $\delta_n\!>\!0$ and $n\geq 1$, there exist an encoder, decoder, and estimators such that (\ref{eq:rates_cons})-(\ref{eq:distortion_consts}) are satisfied for some tuple $(R_1,R_2,D_1,D_2)$. Using Fano's inequality and (\ref{eq:reliability_cons}), we have
  \begin{align}
    H(M|Y_1^n,S_1^n)\!\overset{(a)}{\leq}\!H(M|\widehat{M})\!\leq\!n\epsilon_n \label{eq:fanoapp} 
  \end{align}
  where $(a)$ allows randomized decoding and $\epsilon_n\!=\!\delta_n (R_1\!+\!R_2)\!+\!H_b(\delta_n)/n$ such that $\epsilon_n\!\rightarrow\!0$ if $\delta_n\!\rightarrow\!0$. 
  
  Let $V_{i}\triangleq (M_1,M_2,Y^{i-1}_{1},S_1^{i-1},Y^{i-1}_{2},S_2^{i-1})$ such that $V_i-X_i-(Y_{1,i},Y_{2,i},S_{1,i},S_{2,i})$ form a Markov chain for all $i\in[1:~n]$ by definition of the channel statistics.
  
  \textbf{Bound on $\mathbf{R_1}$}: We have
  \begin{align}
    &nR_1 \overset{(a)}{\leq} I(M_1; Y_1^n|S_1^n)+n\epsilon_n\nonumber\\
    &\leq \sum_{i=1}^n \big(H(Y_{1,i}|S_{1,i})-H(Y_{1,i}|M_1,M_2,Y_1^{i-1},S_1^n)+\epsilon_n\big)\nonumber
                \end{align}
\begin{align}
    & \overset{(b)}{\leq}\sum_{i=1}^n \big(H(Y_{1,i}|S_{1,i})\nonumber\\
    &\qquad\qquad -H(Y_{1,i}|M_1,M_2,Y_{1}^{i-1},S_{1}^{i},Y_2^{i-1},S_2^{i-1})+\epsilon_n\big)\nonumber\\
    &\overset{(c)}{=}\sum_{i=1}^n \big(I(V_i;Y_{1,i}|S_{1,i}) +\epsilon_n)\label{eq:boundonR1converse}
  \end{align}
  where $(a)$ follows by (\ref{eq:fanoapp}) and because $M_1$ and $S_1^n$ are independent, $(b)$ follows since
  \begin{align}
    S_{1,i+1}^{n}-(M_1,M_2,Y_1^{i-1},S_{1}^{i})-Y_{1,i}\label{eq:Markovsi+1}
  \end{align}
  form a Markov chain, and $(c)$ follows from the definition of $V_i$.
  
  \textbf{Bound on $\mathbf{(R_1+R_2)}$}: Similar to (\ref{eq:boundonR1converse}), we obtain
  \begin{align}
    &n(R_1+R_2)\overset{(a)}{\leq} I(M_1,M_2;Y_1^n|S_1^n) +n\epsilon_n\nonumber\\
    &\overset{(b)}{\leq}\sum_{i=1}^n \big(H(Y_{1,i}|S_{1,i})\nonumber\\
    &\qquad\qquad -H(Y_{1,i}|M_1,M_2,Y_{1}^{i-1},S_{1}^{i},Y^{i-1}_{2},S_2^{i-1})+\epsilon_n\big)\nonumber\\
    &\overset{(c)}{=}\sum_{i=1}^n \big(I(V_i;Y_{1,i}|S_{1,i}) +\epsilon_n)
  \end{align}
  where $(a)$ follows because $(M_1,M_2,S_1^n)$ are mutually independent and by (\ref{eq:fanoapp}),  $(b)$ follows since (\ref{eq:Markovsi+1}) form a Markov chain, and $(c)$ follows from the definition of $V_i$.
  
  \textbf{Bound on $\mathbf{R_2}$}: We obtain
  \begin{align}
    &nR_2 \overset{(a)}{\leq}I(M_2;Y_1^n,Y_2^n,S_1^n,S_2^n)+n\epsilon_n\nonumber\\
    &\leq H(Y_{1}^n,S_{1}^n|Y_{2}^n,S_{2}^n)+H(Y_2^n,S_2^n) -H(Y_2^n,S_2^n|M_2)\nonumber\\
    &\qquad -H(Y_1^n,S_1^n|Y_2^n,S_2^n,M_1,M_2)+n\epsilon_n\nonumber\\
    &\leq H(Y_{1}^n,S_{1}^n|Y_{2}^n,S_{2}^n)+I(Y_2^n,S_2^n;M_2)\nonumber\\
    &\qquad  -\sum_{i=1}^nH(S_{1,i}|Y_1^n,Y_2^n,S_2^n,M_1,M_2,S_1^{i-1})+n\epsilon_n\nonumber\\
    &\overset{(b)}{\leq}H(Y_{1}^n,S_{1}^n|Y_{2}^n,S_{2}^n)+\delta_n \nonumber\\
    &\qquad -\sum_{i=1}^nH(S_{1,i}|Y_1^i,Y_2^i,S_2^i,M_1,M_2,S_1^{i-1})+n\epsilon_n\nonumber\\
    &\overset{(c)}{=}H(Y_{1}^n,S_{1}^n|Y_{2}^n,S_{2}^n)+\delta_n \nonumber\\
    &\qquad -\sum_{i=1}^nH(S_{1,i}|Y_{1,i},Y_{2,i},S_{2,i},V_i)+n\epsilon_n\nonumber\\
    &\leq\sum_{i=1}^n \big(H(Y_{1,i},S_{1,i}|Y_{2,i},S_{2,i})- H(S_{1,i}|Y_{1,i},Y_{2,i},S_{2,i},V_i))\big)\nonumber\\
    &\qquad +n\epsilon_n+\delta_n
  \end{align}
  where $(a)$ follows by (\ref{eq:fanoapp}), $(b)$ follows by (\ref{eq:secrecyleakage_cons}) and from Remark~\ref{rem:secrecyconstraintwithoutcond}, and because 
  \begin{align}
    (Y_{1,i+1}^{n},Y_{2,i+1}^{n},S_{2,i+1}^{n})- (Y_{1}^i,Y_{2}^i,S_{2}^i,M_1,M_2,S_1^{i-1})-S_{1,i}
  \end{align}
  form a Markov chain, and $(c)$ follows from the definition of $V_i$.
  
  \textbf{Distortion Bounds}: We have for $j=1,2$ 
  \begin{align}
    (D_j\!+\!\delta_n)\overset{(a)}{\geq} \mathbb{E}\big[d_j(S_j^n,\widehat{S_j^n})\big] = \frac{1}{n}\sum_{i=1}^n\mathbb{E}\big[d_j(S_{j,i},\widehat{S_{j,i}})\big] \label{eq:outerbounddistortion}
  \end{align}
  where $(a)$ follows by (\ref{eq:distortion_consts}), which can be achieved by using the deterministic per-letter estimators in (\ref{eq:deterministicest_new}).
  
  Introduce a uniformly distributed time-sharing random variable $\displaystyle Q\!\sim\! \text{Unif}[1\!:\!n]$ that is independent of other random variables, and define $Y_1\!=\!Y_{1,Q}$, $\displaystyle S_1\!=\!S_{1,Q}$, $\displaystyle Y_2\!=\!Y_{2,Q}$, $\displaystyle S_2\!=\!S_{2,Q}$, $X\!=\!X_Q$, and $V\!=\!(V_{Q},\!Q)$, so $V-X-(Y_1,Y_2,S_1,S_2)$ form a Markov chain. The proof of the outer bound follows by letting $\delta_n\rightarrow0$.
  
  \textbf{Cardinality Bounds}: We use the support lemma \cite[Lemma 15.4]{CsiszarKornerbook2011} to prove the cardinality bound, which is a standard procedure, so we omit the proof.
\end{IEEEproof}

\subsection{Degraded and Reversely-Degraded Channels Under Partial Secrecy}

We next characterize the exact secrecy-distortion regions for physically-degraded and reversely-physically-degraded ISAC channels, which are defined below.

\begin{definition}\label{def:physicallydegraded}
  \normalfont An ISAC channel $P_{Y_1Y_2|S_1S_2X}$ is \emph{physically-degraded} if we have
  \begin{align}
    &P_{Y_1Y_2S_1S_2|X}=P_{Y_1Y_2|S_1S_2X}P_{S_1S_2}\nonumber\\
    &=P_{S_1}P_{Y_1|S_1X}P_{Y_2S_2|S_1Y_1}\label{eq:physicaldegradedcond}
  \end{align}
  and is \emph{reversely-physically-degraded} if the degradation order is changed such that
  \begin{align}
    &P_{Y_1Y_2S_1S_2|X}=P_{Y_1Y_2|S_1S_2X}P_{S_1S_2}\nonumber\\
    &=P_{S_2}P_{Y_2|S_2X}P_{Y_1S_1|S_2Y_2}.\label{eq:reverselyphysicaldegradedcond}
  \end{align}\hfill $\lozenge$
\end{definition}
A physically-degraded ISAC channel corresponds to a situation in which the observation $Y_2$ of the eavesdropper given its state $S_2$ is a degraded version of the observation $Y_1$ of the legitimate receiver given its state $S_1$ with respect to the channel input $X$. 

\begin{theorem}\label{theo:SimplifiedforPSPOFDegraded}
  {\normalfont(Physically-degraded Channels):} For a physically-degraded ISAC channel, $\mathcal{R}_{\textnormal{PS,POF}}$ is the union over all joint distributions $P_{VX}$ of the rate tuples $(R_{1}, R_{2},D_1,D_2)$ satisfying (\ref{eq:achdistortion1and2_new})-(\ref{eq:convR2}), where we have (\ref{eq:jointprobiid_new}) with constant $U$ and (\ref{eq:deterministicest_new}). One can limit $|\mathcal{V}|$ to (\ref{eq:cardVplus1}).
\end{theorem}

\begin{IEEEproof}[Proof of Theorem~\ref{theo:SimplifiedforPSPOFDegraded}]
  Since the outer bound given in Proposition~\ref{prop:OuterforPSPOF} does not assume any degradedness, the outer bound terms for $R_1$, $R_2$, and $D_j$ for $j=1,2$ follow from Proposition~\ref{prop:OuterforPSPOF}.

  The achievability proof for Theorem~\ref{theo:SimplifiedforPSPOFDegraded} follows by modifying the construction and analysis of Protocol A in the proof of Proposition~\ref{prop:InnerforPSPOF}. We next provide a sketch of the modifications for a physically-degraded ISAC channel. First, $U^n$ is not used, i.e., $U^n$ is eliminated from the achievability proof. Second, to each $v^n$ we assign four random bin indices $(F_{\text{v}},W_{\text{v}_1},W_{\text{v}_2},L_{\text{v}})$ such that $F_{\text{v}}\in[1:2^{n\widetilde{R}_{\text{v}}}]$, $W_{\text{v}_1}\in[1:2^{nR_{\text{v}_1}}]$, $W_{\text{v}_2}\in[1:~2^{nR_{\text{v}_2}}]$, and $L_{\text{v}}\in[1:2^{n\overline{R}_{\text{v}}}]$ independently such that $M_1=W_{\text{v}_1}$ and $M_2=(W_{\text{v}_2},L_{\text{v}})$. As in (\ref{eq:reconstrV}), we impose the reliability constraint
  \begin{align} 
    \widetilde{R}_{\text{v}} > H(V|Y_1,S_1)\label{eq:theo2reliability}
  \end{align}
  as in (\ref{eq:independenceofFvWv}) and (\ref{eq:independenceofYindexnewwwwwwww}) we impose the strong secrecy constraints
  \begin{align}
    &R_{\text{v}_2}+\widetilde{R}_{\text{v}}<H(V|Y_2,S_2)\\
    &\overline{R}_{\text{v}}<H(Y_1|Y_2,S_2,V)
  \end{align} 
  and as in (\ref{eq:sumindependence}) we impose the mutual independence and uniformity constraint
  \begin{align}
    R_{\text{v}_1}+R_{\text{v}_2}+\widetilde{R}_{\text{v}} + \overline{R}_{\text{v}}< H(V).\label{eq:theo2independence}
  \end{align}
  
  We remark that we have $H(V|Y_2,S_2)\geq H(V|Y_1,S_1)$ for all physically-degraded ISAC channels, i.e., we obtain
  \begin{align}
    &[I(V;Y_1|S_1)-I(V;Y_2|S_2)]^{+}\nonumber\\
    &\overset{(a)}{=} H(V|Y_2,S_2)-H(V|Y_1,S_1)\label{eq:plussignfordegraded}
  \end{align}
  where $(a)$ follows because $V$ is independent of $(S_1,S_2)$ and since 
  \begin{align}
    V-X-(Y_1,S_1)-(Y_2,S_2)\label{eq:MarkovVXY1S1Y2S2}
  \end{align}
  form a Markov chain for such ISAC channels. Define
  \begin{align}
    &R_{2,\text{deg}}^{\prime} =[I(V;Y_1|S_1)-I(V;Y_2|S_2)]^{+}+H(Y_1|Y_2,S_2,V)\nonumber\\
    &\overset{(a)}{=} H(V|Y_2,S_2)-H(V|Y_1,S_1)+H(Y_1|Y_2,S_2,V)\nonumber\\
    &\overset{(b)}{=}H(Y_1,V|Y_2,S_2)-H(V|Y_1,S_1,Y_2,S_2)\nonumber\\
    &= H(Y_1|Y_2,S_2)+I(V;S_1|Y_1,Y_2,S_2)\nonumber\\
    &= H(Y_1,S_1|Y_2,S_2)- H(S_1|Y_1,Y_2,S_2,V)
  \end{align}
  where $(a)$ follows by (\ref{eq:plussignfordegraded}) and $(b)$ follows from the Markov chain in (\ref{eq:MarkovVXY1S1Y2S2}). 
  
  Applying the Fourier-Motzkin elimination \cite{FMEbook} to (\ref{eq:theo2reliability})-(\ref{eq:theo2independence}), for any $\epsilon>0$ one can achieve 
  \begin{align}
    &R_1= R_{\text{v}_1} =I(V;Y_1,S_1)-2\epsilon=I(V;Y_1|S_1)-2\epsilon\label{eq:R1forTheo2}
  \end{align}
  and for any $R_1$ that is less than or equal to (\ref{eq:R1forTheo2}), one can simultaneously achieve
  \begin{align}
    &R_2 = R_{\text{v}_2} +\overline{R}_{\text{v}} \nonumber\\
    &= \min\{R_{2,\text{deg}}^{\prime} , (I(V;Y_1|S_1)-R_1)\}-3\epsilon.
  \end{align}
  The construction of Protocol B, the analysis of achievable distortions and sufficiency of deterministic estimators, as well as the derandomization and chaining analysis, follow  as in the proof of Proposition~\ref{prop:InnerforPSPOF} and are omitted for brevity.
\end{IEEEproof}

\begin{theorem}\label{theo:reversePSPOF}
  {\normalfont(Reversely-physically-degraded Channels):} For a reversely-physically-degraded ISAC channel, $\mathcal{R}_{\textnormal{PS,POF}}$ is the union over all joint distributions $P_{VX}$ of the rate tuples $(R_{1}, R_{2},D_1,D_2)$ satisfying (\ref{eq:achdistortion1and2_new}), (\ref{eq:achR1degradedbutnowouterProp2}), and 
  \begin{align}
    R_2\leq\min\big\{H(Y_1|Y_2,S_2), \big(I(V;Y_1|S_1)-R_1\big)\big\}   
  \end{align}
  where we have (\ref{eq:jointprobiid_new}) with constant $U$ and (\ref{eq:deterministicest_new}). One can limit $|\mathcal{V}|$ to 
  \begin{align}
    \min\{|\mathcal{X}|,\;|\mathcal{Y}_1|\!\cdot\!|\mathcal{S}_1|,\;|\mathcal{Y}_2|\!\cdot\!|\mathcal{S}_2|\}.\label{eq:Vcardinalityforreversephydegraded}
  \end{align}
  
\end{theorem}

\begin{IEEEproof}[Proof of Theorem~\ref{theo:reversePSPOF}]
  The achievability proof follows from Proposition~\ref{prop:InnerforPSPOF} after elimination of $U$ from its proof, as in the proof for Theorem~\ref{theo:SimplifiedforPSPOFDegraded}. After removal of $U$, by (\ref{eq:achR2}) we have the inner bound
  \begin{align}
    &R_2\overset{(a)}{\leq} \min \big\{H(Y_1|Y_2,S_2,V), \big(I(V;Y_1|S_1)-R_1\big)\big\}\nonumber\\
       &\overset{(b)}{=}   \min \big\{H(Y_1|Y_2,S_2), \big(I(V;Y_1|S_1)-R_1\big)\big\}
  \end{align}
  where $(a)$ follows since $V$ is independent of $(S_1,S_2)$ and because $H(V|Y_1,S_1)\geq H(V|Y_2,S_2)$ for all reversely-physically-degraded ISAC channels because of the Markov chain 
  \begin{align}
    V-X-(Y_2,S_2)-(Y_1,S_1)\label{eq:MarkovVXY2S2Y1S1}
  \end{align}
  and $(b)$ follows also because of the Markov chain in (\ref{eq:MarkovVXY2S2Y1S1}).

  Since the outer bound in Proposition~\ref{prop:OuterforPSPOF} does not assume any degradedness, the outer bound terms for $R_1$ and $D_j$ for $j=1,2$ follow from Proposition~\ref{prop:OuterforPSPOF}. Furthermore, by (\ref{eq:convR2}) we obtain the outer bound
  \begin{align}
    &R_2\overset{(a)}{\leq} \min\big\{\big(H(Y_1,S_1|Y_2,S_2) - H(S_1|Y_1,Y_2,S_2)\big), \nonumber\\
    &\qquad\qquad\qquad \big(I(V;Y_1|S_1)-R_1\big)\big\}  \nonumber\\
       &=   \min \big\{H(Y_1|Y_2,S_2), \big(I(V;Y_1|S_1)-R_1\big)\big\}
  \end{align}
  where $(a)$ follows from the Markov chain in (\ref{eq:MarkovVXY2S2Y1S1}).
\end{IEEEproof}

\section{ISAC Under Full Secrecy}\label{sec:SingleMess}
We next give inner and outer bounds for the situation, in which $M=M_2$ should be kept secret from the eavesdropper and $M_1=\varnothing$. For this situation, the definitions of an achievable secrecy-distortion tuple $(R,D_1,D_2)$ and corresponding strong secrecy-distortion region $\mathcal{R}_{\textnormal{POF}}$ follow from Definition~\ref{def:systemmodel} by eliminating $(M_1,R_1)$ and replacing $(M_2,R_2, \mathcal{R}_{\textnormal{PS,POF}})$ with $(M,R,\mathcal{R}_{\textnormal{POF}})$, respectively.

\begin{proposition}\label{prop:InnerforPOF}
  \emph{(Inner Bound):} The region $\mathcal{R}_{\textnormal{POF}}$ includes the union over all joint distributions $P_{VX}$ of the rate tuples $(R,D_1,D_2)$ satisfying
  \begin{align}
    & D_j\geq \mathbb{E}[d_j(S_j,\widehat{S}_j))]\qquad\qquad  \text{for }j=1,2\label{eq:achdistortion1and2_newfull}\\
    &R\leq \min\{R^{\prime\prime}, I(V;Y_1|S_1)\}\label{eq:achRforPOF}
  \end{align}
  where  
  \begin{align}
    &P_{VXY_1Y_2S_1S_2} = P_{V|X}P_XP_{S_1S_2}P_{Y_1Y_2|S_1S_2X},\label{eq:jointprobiidforPOF}\\
    &R^{\prime\prime}=[I(V;Y_1|S_1)-I(V;Y_2|S_2)]^{+}\nonumber\\
    &\qquad\qquad+H(Y_1|Y_2,S_2,V)\label{eq:R2doubleprime}
  \end{align}
  and one can apply the deterministic per-letter estimators
    \begin{align}
  			&\Est_j(x,y_1,y_2)\nonumber\\
  			&\;=\mathop{\textnormal{argmin}}_{\tilde{s}\in\widehat{\mathcal{S}}_j} \sum_{s_j\in\mathcal{S}_j}P_{S_j|XY_1Y_2}(s_j|x,y_1,y_2)\; d_j(s_j,\tilde{s}).\label{eq:deterministicest_full} 
  \end{align}
   One can limit $|\mathcal{V}|$ to (\ref{eq:cardVplus1}).
\end{proposition}

\begin{IEEEproof}[Proof of Proposition~\ref{prop:InnerforPOF}]
  The proof follows by eliminating $U$ in the proof of Proposition~\ref{prop:InnerforPSPOF}, so $R_1=R_{\text{v}_1}=0$ and by imposing (\ref{eq:theo2reliability})-(\ref{eq:theo2independence}) after replacing $R_{\text{v}_2}$ with $R_{\text{v}}$, since for this case we have $M=(W_{\text{v}},L_{\text{v}})$.
\end{IEEEproof}

\begin{proposition}\label{prop:OuterforPOF}
  \emph{(Outer Bound):} The region $\mathcal{R}_{\textnormal{POF}}$ is included in the union over all $P_X$ of the rate tuples $(R,D_1,D_2)$ satisfying (\ref{eq:achdistortion1and2_newfull}) and
  \begin{align}
    &R\leq \min\Big\{\big(H(Y_1,S_1|Y_2,S_2) - H(S_1|Y_1,Y_2,S_2,X)\big), \nonumber\\
    &\qquad\qquad\qquad I(X;Y_1|S_1)\Big\}    \label{eq:convRforPOF}
  \end{align}
  where one can apply the deterministic per-letter estimators in (\ref{eq:deterministicest_full}). 
\end{proposition}

\begin{IEEEproof}[Proof of Proposition~\ref{prop:OuterforPOF}]
  The proof follows from the proof of Proposition~\ref{prop:OuterforPSPOF} by making appropriate replacements for the full secrecy scenario, but we provide a new proof with minor simplifications for completeness. Assume that for some $\delta_n\!>\!0$ and $n\geq 1$, there exist an encoder, a decoder, and estimators such that all constraints imposed on the ISAC problem with perfect output feedback are satisfied for some tuple $(R,D_1,D_2)$. We then obtain
  \begin{align}
    &nR\overset{(a)}{\leq} I(M;Y_1^n|S_1^n)+n\epsilon_n\nonumber\\
    &\leq \sum_{i=1}^n \big(H(Y_{1,i}|S_{1,i}) - H(Y_{1,i}|Y_1^{i-1},S_1^n,M,X_i)+\epsilon_n\big)\nonumber\\
    &\overset{(b)}{=} \sum_{i=1}^n \big(H(Y_{1,i}|S_{1,i}) - H(Y_{1,i}|S_{1,i},X_i)+\epsilon_n\big)\nonumber\\
    &= \sum_{i=1}^n (I(X_i;Y_{1,i}|S_{1,i})+\epsilon_n)\label{eq:outerboundforsinglemessageIXY1givenS1}
  \end{align}
  where $(a)$ follows because $M$ and $S_1^n$ are independent, and from Fano's inequality for an $\epsilon_n>0$ such that $\epsilon_n\rightarrow 0$ if $\delta_n\rightarrow 0$, which is entirely similar to (\ref{eq:fanoapp}), and $(b)$ follows because 
  \begin{align}
    Y_{1,i}-(S_{1,i},X_i)-(Y_1^{i-1},S_1^{n\setminus i},M)
  \end{align}
  form a Markov chain. Furthermore, we also have
  \begin{align}
    &nR\overset{(a)}{\leq}I(M;Y_1^n,Y_2^n,S_1^n,S_2^n)+n\epsilon_n\nonumber\\
    &= H(Y_1^n,S_1^n|Y_2^n,S_2^n)+I(Y_2^n,S_2^n;M) \nonumber\\
    &\qquad -H(Y_1^n,S_1^n|Y_2^n,S_2^n,M)+n\epsilon_n\nonumber\\
    &\overset{(b)}{\leq} \sum_{i=1}^nH(Y_{1,i},S_{1,i}|Y_{2,i},S_{2,i})+ \delta_n \nonumber\\
    &\qquad -\sum_{i=1}^nH(S_{1,i}|Y_1^n,Y_{2}^n,S_{2}^n,M,S_1^{i-1},X_i)+n\epsilon_n\nonumber\\
    &\overset{(c)}{=}\sum_{i=1}^n\Big(H(Y_{1,i},S_{1,i}|Y_{2,i},S_{2,i})\nonumber\\
    &\qquad\qquad -H(S_{1,i}|Y_{1,i},Y_{2,i},S_{2,i},X_i)+\epsilon_n\Big)+\delta_n
  \end{align}
  where $(a)$ follows from Fano's inequality, which is similar to (\ref{eq:fanoapp}), $(b)$ follows by (\ref{eq:secrecyleakage_cons}) and from Remark~\ref{rem:secrecyconstraintwithoutcond} after replacing $M_2$ with $M$ for the ISAC problem with a single secure message, and $(c)$ follows because 
  \begin{align}
    S_{1,i}-(Y_{1,i},Y_{2,i},S_{2,i},X_i)-(Y_1^{n\setminus i},Y_2^{n\setminus i}, S_2^{n\setminus i},M,S_1^{i-1})\label{eq:Markovsinglesecmessageouternew}
  \end{align}
  form a Markov chain. Thus, by applying the distortion bounds in (\ref{eq:outerbounddistortion}) and introducing a uniformly-distributed time-sharing random variable, as being applied in the proof of Proposition~\ref{prop:OuterforPSPOF}, we prove the outer bound for the ISAC problem with a single secure message and perfect output feedback by letting $\delta_n\rightarrow 0$.
\end{IEEEproof}

\subsection{Degraded and Reversely-Degraded Channels Under Full Secrecy}

We next present the exact strong secrecy-distortion regions for the ISAC problem with a single secure message when the ISAC channel $P_{Y_1Y_2|S_1S_2X}$ is physically-degraded, as in (\ref{eq:physicaldegradedcond}), or reversely-physically-degraded, as in (\ref{eq:reverselyphysicaldegradedcond}).

\begin{theorem}\label{theo:SimplifiedforPOF}
  {\normalfont(Physically-degraded Channels):} For a physically-degraded ISAC channel, $\mathcal{R}_{\textnormal{POF}}$ is the union over all probability distributions $P_{X}$ of the rate tuples $(R,D_1,D_2)$ satisfying (\ref{eq:achdistortion1and2_newfull}) and (\ref{eq:convRforPOF}), where we have (\ref{eq:deterministicest_full}). 
\end{theorem}

\begin{IEEEproof} [Proof of Theorem~\ref{theo:SimplifiedforPOF}]
  Since the bound given in Proposition~\ref{prop:OuterforPOF} is valid for any ISAC channel, the proof for the outer bound follows from Proposition~\ref{prop:OuterforPOF}. Furthermore, the achievability proof follows by modifying the proof of Theorem~\ref{theo:SimplifiedforPSPOFDegraded} such that we assign $V^n(k)=X^n(k)$ for all $k=[1:b]$ and then apply the same OSRB steps for $X^n(k)$ rather than $V^n(k)$, i.e., replace $V$ with $X$ in the inner bound terms given in Proposition~\ref{prop:InnerforPOF}. Define
  \begin{align}
    &R^{\prime\prime}_{\text{deg}}=[I(X;Y_1|S_1)-I(X;Y_2|S_2)]^{+}+H(Y_1|Y_2,S_2,X)\nonumber\\
    &\overset{(a)}{=}I(X;Y_1,S_1|Y_2,S_2)+H(Y_1|Y_2,S_2,X)\nonumber\\
    &=H(Y_1,S_1|Y_2,S_2)-H(S_1|Y_1,Y_2,S_2,X)
  \end{align}
  where $(a)$ follows because the ISAC channel is physically-degraded, and since $X$ is independent of $(S_1,S_2)$. Thus, by (\ref{eq:achRforPOF}) we have
  \begin{align}
    &R\leq \min\{R^{\prime\prime}_{\text{deg}}, I(X;Y_1|S_1)\}
  \end{align}
  which proves the achievability bound.
\end{IEEEproof}

\begin{theorem}\label{theo:reversePOFsimplified}
  {\normalfont(Reversely-physically-degraded Channels):} For a reversely-physically-degraded ISAC channel, $\mathcal{R}_{\textnormal{POF}}$ is the union over all probability distributions $P_{X}$ of the rate tuples $(R,D_1,D_2)$ satisfying (\ref{eq:achdistortion1and2_newfull}) and
  \begin{align}
    R\leq\min\big\{H(Y_1|Y_2,S_2), I(X;Y_1|S_1)\big\}   
  \end{align}
  where one can apply the deterministic per-letter estimators in (\ref{eq:deterministicest_full}).	
\end{theorem}

\begin{IEEEproof}[Proof of Theorem~\ref{theo:reversePOFsimplified}]
  We assign $V^n=X^n$ in the achievability proof, i.e., we choose $V=X$ that is allowed by (\ref{eq:jointprobiidforPOF}), such that by (\ref{eq:achRforPOF}) we obtain the inner bound
  \begin{align}
    &R\overset{(a)}{\leq} \min \big\{H(Y_1|Y_2,S_2,X), I(X;Y_1|S_1)\big\}\nonumber\\
     &\overset{(b)}{=}   \min \big\{H(Y_1|Y_2,S_2), I(X;Y_1|S_1)\big\}
  \end{align}
  where $(a)$ follows since $X$ is independent of $(S_1,S_2)$ and because $H(X|Y_1,S_1)\geq H(X|Y_2,S_2)$ for all reversely-physically-degraded ISAC channels due to the Markov chain in (\ref{eq:MarkovVXY2S2Y1S1}), and $(b)$ follows also because of the Markov chain in (\ref{eq:MarkovVXY2S2Y1S1}).

  Since the outer bound in Proposition~\ref{prop:OuterforPOF} does not assume any degradedness, the outer bound terms for $D_j$ for $j=1,2$ follow from Proposition~\ref{prop:OuterforPOF}. Furthermore, by (\ref{eq:convRforPOF}) we have the outer bound
  \begin{align}
    &R\overset{(a)}{\leq} \min\Big\{\big(H(Y_1,S_1|Y_2,S_2) - H(S_1|Y_1,Y_2,S_2)\big),  \nonumber\\
    &\qquad\qquad\qquad I(X;Y_1|S_1)\Big\}  \nonumber\\
     &=   \min \big\{H(Y_1|Y_2,S_2), I(X;Y_1|S_1)\big\}
  \end{align}
  where $(a)$ follows from the Markov chain in (\ref{eq:MarkovVXY2S2Y1S1}).
\end{IEEEproof}

\section{ISAC Channels with Bernoulli States}\label{sec:JCASexamples}
\subsection{Binary Noiseless ISAC Channels with Bernoulli States}\label{subsec:JCASexample1}
We next consider a scenario with perfect output feedback and single secure message, in which channel input and output alphabets are binary with multiplicative Bernoulli states, which serves as a coarse model of fading channels with high signal-to-noise ratio. Specifically, we have
\begin{align}
  Y_1 =S_1\cdot X,\qquad\qquad\qquad Y_2=S_2\cdot X
\end{align}
and
\begin{align}
  &P_{S_1S_2}(0,0)\!=\!(1\!-\!q), \qquad
  P_{S_1S_2}(1,1)\!=\!q\alpha,\nonumber\\
  & P_{S_1S_2}(0,1)\!=\!0,\qquad\qquad
  P_{S_1S_2}(1,0)\!=\!q(1\!-\!\alpha)\label{eq:PS1S2exampledistribution} 
\end{align}
for fixed $q,\alpha\in[0,1]$, so the ISAC channel is stochastically-degraded, i.e., there exists a marginal probability distribution such that the ISAC channel can be represented as in (\ref{eq:physicaldegradedcond}). The constraints (\ref{eq:rates_cons})-(\ref{eq:distortion_consts}) in Definition~\ref{def:systemmodel} only depend on the marginal probability distributions of $(X,Y_1,S_1)$ and $(X,Y_2,S_2)$ when per-letter estimators of the form $\Est_j(x,y_j)$ are imposed for $j=1,2$, so the secrecy-distortion region given in Theorem~\ref{theo:SimplifiedforPOF} is also valid for stochastically-degraded ISAC channels.

\begin{lemma}\label{lem:firstexampleregion}
  The strong secrecy-distortion region $\mathcal{R}_{\textnormal{POF}}$ for a binary ISAC channel with multiplicative Bernoulli states characterized by parameters $(q,\alpha)$ and with Hamming distortion metrics is the union over all $p\in[0,1]$, where $X\sim\text{Bern}(p)$, of the rate tuples  $(R,D_1,D_2)$ satisfying
  \begin{align}
    &R\leq \min\Bigg\{\Bigg(q(1-\alpha)H_b(p)+ p(1-q\alpha)H_b\Big(\frac{q(1-\alpha)}{(1-q\alpha)}\Big)\Bigg), \nonumber\\
    &\qquad\qquad\qquad\qquad  qH_b(p)\Bigg\}\label{eq:rateforexample}\\
    &D_1\geq (1-p)\cdot \min\{q,\; (1-q)\}\label{eq:distortion1forexample}\\
    &D_2\geq (1-p)\cdot\min\{q\alpha,\; (1-q\alpha)\}.\label{eq:distortion2forexample}
  \end{align}
\end{lemma}

\begin{IEEEproof}[Proof of Lemma~\ref{lem:firstexampleregion}] 
  The proof follows by evaluating the strong secrecy-distortion region $\mathcal{R}_{\textnormal{POF}}$ defined in Theorem~\ref{theo:SimplifiedforPOF}. Proofs for (\ref{eq:distortion1forexample}) and (\ref{eq:distortion2forexample}) follow by choosing $\Est_j(1,y_j)=y_j$ and $\Est_j(0,y_j)=\mathds{1}\{\Pr[S_j=1]>0.5\}$ for $j=1,2$ that can be obtained as in (\ref{eq:deterministicest_full}), which are equivalent to the proofs for  \cite[Eqs. (27c) and (27d)]{MariMicheleBCJCAS}. We next have $I(X;Y_1|S_1)=qH_b(p)$, which is equivalent to the proof for \cite[Eq. (27a)]{MariMicheleBCJCAS} with $r=1$. Furthermore, we obtain
  \begin{align}
    &H(Y_1,S_1|Y_2,S_2)-H(S_1|Y_1,Y_2,S_2,X)\nonumber\\
    &\overset{(a)}{=}H(S_1|S_2)+H(Y_1|S_1,Y_2,S_2)-H(S_1|S_2)\nonumber\\
    &\qquad +I(S_1;Y_1,X|S_2)\nonumber\\
    &\overset{(b)}{=}P_{S_1S_2}(1,0)H(Y_1|S_1=1,S_2=0)+H(X)\nonumber\\
    &\qquad +H(Y_1|X,S_2)-H(Y_1,X|S_2,S_1)\nonumber\\
    &\overset{(c)}{=}P_{S_1S_2}(1,0)H(X)+H(X)\nonumber\\
    &\qquad +P_{X}(1)P_{S_2}(0)H(Y_1|X=1,S_2=0)\nonumber\\
    &\qquad +P_X(1)P_{S_2}(1)H(Y_1|X=1,S_2=1)-H(X)\nonumber\\
    &\overset{(d)}{=}q(1-\alpha)H_b(p)+ p(1-q\alpha)H_b\Big(\frac{q(1-\alpha)}{(1-q\alpha)}\Big)
  \end{align}
  where $(a)$ follows since $S_1-S_2-Y_2$ and $S_1-(Y_1,S_2,X)-Y_2$ form Markov chains for the considered ISAC channel, $(b)$ follows since if $S_1=0$, then $Y_1=0$; if $(S_1,S_2)=(1,1)$, then $Y_1=Y_2=X$; and if $S_2=0$, then $Y_2=0$, and because $X$ is independent of $S_2$, $(c)$ follows since $Y_1=X$ if $S_1=~1$, because $X$ is independent of $(S_1,S_2)$, since $Y_1=0$ if $X=~0$, and because $(S_1,X)$ determine $Y_1$, and $(d)$ follows since $S_1=~1$ if $S_2=1$ due to (\ref{eq:PS1S2exampledistribution}) and because $(S_1,X)$ determine $Y_1$. Therefore, we have
  \begin{align}
    &R\leq \min\Big\{\big(H(Y_1,S_1|Y_2,S_2) - H(S_1|Y_1,Y_2,S_2,X)\big), \nonumber\\
    &\qquad\qquad\qquad I(X;Y_1|S_1)\Big\}\nonumber\\
     &= \min\Bigg\{\Bigg(q(1-\alpha)H_b(p)+ p(1-q\alpha)H_b\Big(\frac{q(1-\alpha)}{(1-q\alpha)}\Big)\Bigg),  \nonumber\\
     &\qquad\qquad\qquad qH_b(p)\Bigg\}.
  \end{align}
\end{IEEEproof}

\begin{figure}[t]
  \centering
  \input{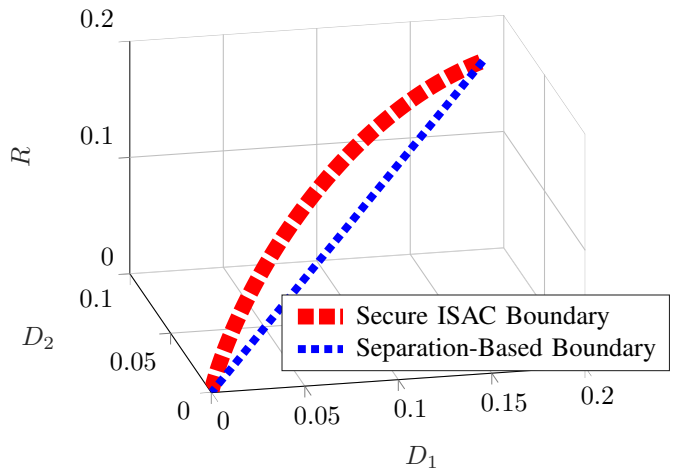}
  \caption{Boundary of the secrecy-distortion region $\mathcal{R}_{\textnormal{POF}}$ for a binary ISAC channel with multiplicative Bernoulli states characterized by parameters $(q=~0.65,\alpha=0.21)$ and with Hamming distortion metrics, as well as the separation-based region boundary.} 
  \label{fig:illustrattiveplot}
  \vspace*{-0.3cm}
\end{figure}

The securely-transmitted message rate for ISAC scenarios under full secrecy is upper bounded both by $\big(H(Y_1,S_1|Y_2,S_2) -H(S_1|Y_1,Y_2,S_2,X)\big)$ and $I(X;Y_1|S_1)$, the latter of which is the upper bound for the rate when there is no secrecy constraint \cite[Corollary~4]{MariMicheleBCJCAS}. Thus, secrecy might incur a rate penalty for this example. Nevertheless, ISAC methods achieve significantly better performance than separation-based secure communication and state-sensing methods. 
One can illustrate this by showing that time sharing between the operation point with the maximum secrecy rate and the point with the minimum distortions results in a region that is strictly smaller than the one identified in Lemma~\ref{lem:firstexampleregion}; see Fig.~\ref{fig:illustrattiveplot} for the boundary of the secrecy-distortion region $\mathcal{R}_{\textnormal{POF}}$ for a binary ISAC channel with multiplicative Bernoulli states characterized by parameters $(q=0.65,\alpha=0.21)$. These analyses are analogous to the comparisons between joint and separation-based secrecy and reliability methods for the secret key agreement problem, as discussed in \cite{bizimWZ,Blochpaper,bizimPeterISITA}.

\subsection{BEC-BSC ISAC Channels with State-Dependent Inputs}\label{subsec:JCASexample2}
We next illustrate an achievable rate region for a noisy ISAC channel. Consider a binary-input ISAC channel with binary channel states such that 
\begin{align}
  &X\sim \text{Bern}(p),\label{eq:BECBSCPX}\\
  &\widebar{X}_1=S_1\cdot X,\qquad \qquad \widebar{X}_2=S_2\cdot X,\label{eq:BECBSCXbars}\\
  &P_{Y_1|\widebar{X}_1}\sim \text{BEC}(\gamma),\label{eq:BECBSCBEC}\\ &P_{Y_2|\widebar{X}_2} \sim \text{BSC}(\beta)\label{eq:BECBSCBSC}
\end{align}
for $ \gamma, \beta \in (0,1)$, $p\in[0,1]$, and the binary states $(S_1,S_2)$ are again distributed according to the joint probability distribution in (\ref{eq:PS1S2exampledistribution}). We remark that (\ref{eq:BECBSCBEC}) and (\ref{eq:BECBSCBSC}) impose the following Markov chains
\begin{align}
  &Y_1 - (S_1,X) - (Y_2,S_2),\label{eq:MArkovY1S1XY2S2}\\
  &Y_2 - (S_2,X) - (Y_1,S_1).\label{eq:MArkovY2S2XY1S1}
\end{align}
We next establish the set of $(\gamma,\beta)$ parameters such that the BEC-BSC ISAC channel is more-capable, as defined below.

\begin{definition}\label{def:morecapable}
  \normalfont An ISAC channel $P_{Y_1Y_2|S_1S_2X}$ is \emph{more-capable} if we have for all $P_X$
  \begin{align}
    I(X;Y_1,S_1)\geq I(X;Y_2,S_2) \label{eq:MCdef}.
  \end{align}\hfill$\lozenge$
\end{definition}

The set of more-capable channels is strictly larger than the set of degraded channels. Furthermore, (\ref{eq:MCdef}) is equivalent to $I(X;Y_1|S_1)\geq I(X;Y_2|S_2)$ since $X$ is independent of $(S_1,S_2)$ for ISAC models. 


\begin{lemma}\label{lem:BECBSCEx}
  BEC-BSC ISAC channels with state-dependent inputs, as defined in (\ref{eq:BECBSCPX})-(\ref{eq:BECBSCBSC}), are more-capable if we have
  \begin{align}
    \gamma\leq 1- \alpha(1-H_b(\beta)).\label{eq:morecapableBECBSCcondition}
  \end{align}
\end{lemma}

We present the proof of Lemma~\ref{lem:BECBSCEx} in Appendix.

We next evaluate an inner bound for the strong secrecy-distortion region of more-capable BEC-BSC ISAC channels. For simplicity in the bound below for $D_2$, suppose $\beta\in (0,0.5]$. The results for $\beta\in (0.5, 1)$ follow by symmetry.

\begin{lemma}\label{lem:MCBECBSCJCAS}
  For more-capable BEC-BSC ISAC channels with state-dependent inputs, defined in (\ref{eq:BECBSCPX})-(\ref{eq:BECBSCBSC}) for $ \gamma \in (0,1)$ and $\beta\in (0, 0.5]$, satisfying (\ref{eq:morecapableBECBSCcondition}), and with Hamming distortion metrics, $\mathcal{R}_{\textnormal{POF}}$ includes the union over all $p\in[0,1]$ of the rate tuples $(R,D_1,D_2)$ satisfying 
  \begin{align}
    &R\leq q\big(H_b(p)(1-\gamma)+\alpha(H_b(\beta)-H_b(p*\beta))\big) \label{eq:example2corRbound}\\
    &D_1\geq (1-p+p\gamma)\cdot\min\{q,\; (1-q)\}\\
    &D_2\geq (1-p)\cdot\min\{q\alpha,\;(1-q\alpha)\} \nonumber\\
    &\qquad\quad + p\cdot
         \begin{cases}
           q\alpha\quad &\text{if}\quad q\alpha\leq \beta\\
           (q\alpha\!*\!\beta\!*\!q\alpha)\quad &\text{if}\quad \beta< q\alpha\leq (1-\beta)\\
           (1-q\alpha)\qquad &\text{if}\quad q\alpha>(1-\beta).\label{eq:example2corD2bound}
         \end{cases}
  \end{align}
\end{lemma}

\begin{IEEEproof}[Proof of Lemma~\ref{lem:MCBECBSCJCAS}]
  Consider the inner bound given in Proposition~\ref{prop:InnerforPOF} that is valid for all ISAC channels with single secure message and perfect output feedback. Choose $V=X$ in (\ref{eq:achRforPOF}), which is allowed by (\ref{eq:jointprobiidforPOF}), so we obtain
  \begin{align}
    &R\!\leq \!\min \{\!\big([I(X;Y_1|S_1)\!-\!I(X;Y_2|S_2)]^+ \!+\! H(Y_1|Y_2,S_2,X)\big),\nonumber\\
    &\qquad\qquad\qquad  I(X;Y_1|S_1)\}\nonumber\\
     &\overset{(a)}{=}\min \{\big(I(X;Y_1|S_1)-I(X;Y_2|S_2) + H(Y_1|S_2,X)\big),\nonumber\\
     &\qquad\qquad\qquad I(X;Y_1|S_1)\}\label{eq:boundonRforsecondexample}
  \end{align}
  where $(a)$ follows by (\ref{eq:MCdef}) since the BEC-BSC ISAC channel is more-capable and from the Markov chain in (\ref{eq:MArkovY2S2XY1S1}). Using (\ref{eq:Dpformula}), we obtain 
  \begin{align}
    &(I(X;Y_1|S_1)-I(X;Y_2|S_2))\nonumber\\
    &  = q\big(H_b(p)(1-\gamma)+\alpha(H_b(\beta)-H_b(p*\beta))\big). 
  \end{align}
  We remark that the calculation of the term $H(Y_1|S_2,X)$ in (\ref{eq:boundonRforsecondexample}) is cumbersome, so we omit it for simplicity since the rate region given in Lemma~\ref{lem:MCBECBSCJCAS} is an inner bound. We next calculate the achievable distortions.

  Choose the first estimator as
  \begin{align*}
    &\Est_1(x,y_1)\nonumber\\
    &=
    \begin{cases}
      \mathds{1}\{\Pr[S_1=1]>0.5\}\quad &\text{if}\quad (x,y_1)=(0,0)\text{ or } (\cdot,\mathtt{e})\\
      y_1\quad &\text{if}\quad (x,y_1)=(1,1)\text{ or } (1,0)
    \end{cases}
  \end{align*}
  which minimizes the probability of error given $(x,y_1)$. Thus, by using (\ref{eq:achdistortion1and2_newfull}) we obtain
  \begin{align}
    &D_1\geq  \mathbb{E}\big[d(S_1,\Est_1(X,Y_1))\big] \nonumber\\
    &\overset{(a)}{=}(P_X(0)+P_{XY_1}(1,\mathtt{e}))\cdot\mathbb{E}\big[d(S_1,\mathds{1}\{\Pr[S_1=1]>0.5\})\big] \nonumber\\
    & \overset{(b)}{=} (1-p+p\gamma)\nonumber\\
    &\qquad\cdot \Big(P_{S_1}(1)\cdot(1\oplus\mathds{1}\{\Pr[S_1=1]>0.5\})\nonumber\\
    &\qquad\qquad\qquad+P_{S_1}(0)\cdot(0\oplus\mathds{1}\{\Pr[S_1=1]>0.5\})\Big) \nonumber\\
    &= (1-p+p\gamma)\cdot \min\{q,\; (1-q)\}\label{eq:D1boundforexample2}
  \end{align}
  where $(a)$ follows since $P_{XY_1}(0,1)=0$ and because there is no estimation error in other cases, and $(b)$ follows because we consider a Hamming distortion metric.
  
  Choose the second estimator as
  \begin{align*}
    &\Est_2(x,y_2)\nonumber\\
    &=
    \begin{cases}
      \mathds{1}\{\Pr[S_2=1]>0.5\}\quad &\text{if}\quad x=0\\
      \mathds{1}\{\Pr[S_2=1|Y_2=y_2,X=x]>0.5\}\quad &\text{if}\quad x=1
    \end{cases}
  \end{align*}
  which minimizes the probability of error given $(x,y_2)$. One can show that
  \begin{align}
    &\mathds{1}\{\Pr[S_2\!=\!1|Y_2\!=\!1,X\!=\!1]\!>\!0.5\} \!=\! \mathds{1}\{q\alpha>\beta\},\label{eq:indicatorfuncev1}\\
    &\mathds{1}\{\Pr[S_2\!=\!1|Y_2\!=\!0,X \! = \!1]\!>\!0.5\} \!=\! \mathds{1}\{q\alpha>(1\!-\!\beta)\}\label{eq:indicatorfuncev2}.
  \end{align}
  Thus, by using (\ref{eq:achdistortion1and2_newfull}), (\ref{eq:indicatorfuncev1}), and (\ref{eq:indicatorfuncev2}), we obtain
  \begin{align}
    &D_2\geq  \mathbb{E}\big[d(S_2,\Est_2(X,Y_2))\big] \nonumber\\
       &=P_X(0)\cdot\mathbb{E}\big[d(S_2,\mathds{1}\{q\alpha>0.5\})\big] \nonumber\\
       &\qquad+ P_{XY_2}(1,1)\cdot \mathbb{E}\big[d(S_2,\mathds{1}\{q\alpha>\beta\})\big] \nonumber\\
       &\qquad + P_{XY_2}(1,0)\cdot \mathbb{E}\big[d(S_2,\mathds{1}\{q\alpha>(1-\beta)\})\big] \nonumber\\
       &\overset{(a)}{=}(1-p)\cdot\min\{q\alpha,\; (1-q\alpha)\}\nonumber\\
       &\qquad + p(q\alpha*\beta)\cdot \mathbb{E}\big[d(S_2,\mathds{1}\{q\alpha>\beta\})\big] \nonumber\\
       &\qquad + p(1-q\alpha*\beta)\cdot \mathbb{E}\big[d(S_2,\mathds{1}\{q\alpha>(1-\beta)\})\big] \label{eq:D2boundforexample2first}
  \end{align}
  where $(a)$ follows by applying a similar step to (\ref{eq:D1boundforexample2})$(b)$. Furthermore, we have
  \begin{align}
    &\mathbb{E}\big[d(S_2,\mathds{1}\{q\alpha>\beta\})\big] = 
      \begin{cases}
        q\alpha\quad&\text{if}\quad q\alpha\leq \beta\\
        (1-q\alpha)\quad&\text{if}\quad q\alpha> \beta,
      \end{cases}\\
    & \mathbb{E}\big[d(S_2,\mathds{1}\{q\alpha>(1-\beta)\})\big]\nonumber\\
    & = 
      \begin{cases}
        q\alpha\quad&\text{if}\quad q\alpha\leq (1-\beta)\\
        (1-q\alpha)\quad&\text{if}\quad q\alpha> (1-\beta).
      \end{cases}\label{eq:expecteddistortionforY2zero}
  \end{align}
  Therefore, since we assume that $\beta\in (0, 0.5]$, i.e., we have $\beta\leq (1-\beta)$, using (\ref{eq:D2boundforexample2first})-(\ref{eq:expecteddistortionforY2zero}) we have
  \begin{align}
    &D_2 \geq (1-p)\cdot\min\{q\alpha,\; (1-q\alpha)\}\nonumber\\
    &\qquad\;\;\;+ p\cdot
      \begin{cases}
        &(q\alpha*\beta)q\alpha +(1-q\alpha*\beta)q\alpha\nonumber\\
        &\qquad\qquad\qquad\qquad \text{if}\quad q\alpha\leq\beta,\\
        &(q\alpha*\beta)(1-q\alpha)+ (1-q\alpha*\beta)q\alpha\nonumber\\
        &\qquad\qquad\qquad\qquad \text{if}\quad \beta<q\alpha\leq(1-\beta),\\
        &(q\alpha*\beta)(1-q\alpha)+ (1-q\alpha*\beta)(1-q\alpha)\nonumber\\
        &\qquad\qquad\qquad\qquad \text{if}\quad q\alpha>(1-\beta)
      \end{cases}
  \end{align}
  which is equal to (\ref{eq:example2corD2bound}).
\end{IEEEproof}


\bibliographystyle{IEEEtran}
\bibliography{references}

\appendix

\begin{IEEEproof}[Proof of Lemma~\ref{lem:BECBSCEx}]
  We follow steps similar to the proofs of \cite[Claims~3 and 4]{NairEssentially}. Define
  \begin{align}
    &D(p) = I(X;Y_1|S_1)-I(X;Y_2|S_2)\nonumber\\
    &\overset{(a)}{=} H_b(p)-H(X|Y_1,S_1)\nonumber\\
    &\qquad\quad-H(Y_2|S_2)+H(Y_2|S_2,X)\label{eq:Dpfirst}
  \end{align}
  where $(a)$ follows since $X\sim\text{Bern}(p)$ is independent of $S_1$. 
  
  First, we have
  \begin{align}
   &H(X|Y_1,S_1)  \overset{(a)}{=}H_b(p)\big(P_{Y_1S_1}(0,0)+P_{Y_1}(\mathtt{e})\big) \nonumber\\
    &= H_b(p)\big((1-q)(1-\gamma)+\gamma\big) \nonumber\\
    &= H_b(p)(1-q+q\gamma)\label{eq:example2HXgivenY1S1}
  \end{align}
  where $(a)$ follows since $H(X|Y_1=0,S_1=1)$ is equal to $H(X|Y_1=1,S_1=1)=0$ and $P_{Y_1S_1}(1,0)=0$.
  
  Second, we have
  \begin{align}
    &H(Y_2|S_2)=P_{S_2}(0)H(Y_2|S_2=0)+P_{S_2}(1)H(Y_2|S_2=1)\nonumber\\
    &\overset{(a)}{=}(1-q\alpha)H_b(\beta)+q\alpha H_b(p*\beta)\label{example2HY2givenS2}
  \end{align}
  where $(a)$ follows since we obtain
  \begin{align*}
    &P_{Y_2|S_2}(0|1)=P_X(0)P_{Y_2|\widebar{X}_2}(0|0)+P_X(1)P_{Y_2|\widebar{X}_2}(0|1)\nonumber\\
    &=p*(1-\beta)
  \end{align*}
  and because the binary entropy function is symmetric around $0.5$.
  
  Third, we have
  \begin{align}
    H(Y_2|S_2,X)= H_b(\beta).\label{eq:HY2givenS2X}
  \end{align}
  Therefore, by combining (\ref{eq:Dpfirst})-(\ref{eq:HY2givenS2X}), have
  \begin{align}
    D(p) = q\big(H_b(p)(1-\gamma)+\alpha(H_b(\beta)-H_b(p*\beta))\big).\label{eq:Dpformula}
  \end{align}
  We next establish the set of $(\gamma,\beta)$ parameters such that the BEC-BSC ISAC channel is more-capable by finding the set of parameters such that $D(p)\geq 0$ for all $p\in[0,1]$. 
  
  Since $D(p)$ given in (\ref{eq:Dpformula}) is symmetric around $0.5$, which follows from the symmetry of the binary entropy function around $0.5$, we can consider the range $p\in[0, 0.5]$ rather than $p\in[0, 1]$. Similarly, we can consider the range $\beta\in (0, 0.5]$ due to symmetry. One can show that $D(0)=0$, and $D(0.5)\geq 0$ if 
  \begin{align}
    \gamma\leq 1-\alpha(1-H_b(\beta)).\label{eq:gammaforDonehalftobepositive}
  \end{align}
  Moreover, we have
  \begin{align}
    &\frac{d}{dp}D(p) \nonumber\\
    &=q\Big(\log\Big(\frac{1-p}{p}\Big)(1-\gamma) - \alpha\log\Big(\frac{1-p*\beta}{p*\beta}\Big)(1-2\beta)\Big)\label{eq:derivativeDP}
  \end{align}
  that is non-negative for all $p\in[0, 0.5]$ if 
  \begin{align}
    \gamma \leq 1-\alpha(1-2\beta)\label{eq:simplergammabound}
  \end{align}
  which follows because $\log((1-p)/p)$ is non-negative and $p*~\beta\geq p$ for all $p\in[0, 0.5]$ and any $\beta\in (0, 0.5]$. Thus, if the parameters satisfy (\ref{eq:simplergammabound}), $D(p)$ is non-decreasing and is non-negative for $p\in [0, 0.5]$ since $D(0)=0$, which proves that the BEC-BSC ISAC channel is more-capable for the parameters that satisfy (\ref{eq:simplergammabound}).
  
  Next, we consider the case 
  \begin{align}
    \gamma > 1-\alpha(1-2\beta)\label{eq:complementofthesimplergammabound}.
  \end{align}
  Define
  \begin{align}
    c = \frac{1-\gamma}{\alpha(1-2\beta)}
  \end{align}
  such that using (\ref{eq:derivativeDP}) we obtain
  \begin{align}
    &\frac{d}{dp}D(p)\geq 0\iff  \Big(\frac{1-p}{p}\Big)^{c}\geq \Big(\frac{1-p*\beta}{p*\beta}\Big)\nonumber\\
    & \iff p*\beta \geq \Bigg(\Big(\frac{1-p}{p}\Big)^c+1\Bigg)^{-1}\label{eq:pstarbetaineq}
  \end{align}
  with equality if $p=0.5$. We remark that $0<c<1$ when we consider (\ref{eq:complementofthesimplergammabound}), which does not allow $\alpha=0$ or $\beta=0.5$ since $\gamma\in (0,1)$. Furthermore, the function 
  \begin{align}
    \Bigg(\Big(\frac{1-p}{p}\Big)^c+1\Bigg)^{-1}
  \end{align}
  is proved in \cite[pp. 10]{NairEssentially} to be concave in $p$ in the range $[0, 0.5]$ if $0< c<1$, which indicates that there can be at most two values of $p$ that achieves equality in (\ref{eq:pstarbetaineq}). Since equality is achieved when $p=0.5$, the other possible value of $p$ that achieves equality in (\ref{eq:pstarbetaineq}) must be in the range $(0,0.5)$. Note that we have
  \begin{align}
    \lim_{p\rightarrow 0^+}\,\frac{d}{dp}D(p) = \infty
  \end{align}
  so we can conclude that $D(p)$ first increases and then decreases in $p$ in the range $p\in[0, 0.5]$ for the parameters that satisfy (\ref{eq:complementofthesimplergammabound}). Combining the conditions in (\ref{eq:gammaforDonehalftobepositive}) and (\ref{eq:complementofthesimplergammabound}), as well as the results for the condition in (\ref{eq:simplergammabound}), the proof follows.
\end{IEEEproof}

\end{document}